\pdfoutput=1
\documentclass[twoside,twocolumn]{article}
\usepackage{blindtext} 

\usepackage[sc]{mathpazo} 
\usepackage[T1]{fontenc} 
\usepackage{microtype} 

\usepackage{graphicx}
\usepackage{amsfonts}
\usepackage{amsmath,amssymb}
\usepackage{multirow}
\usepackage{algorithmic}
\usepackage[linesnumbered, ruled]{algorithm2e}
\usepackage{subfigure}
\newcommand{\ourmodel}{\textbf{A2LH}}
\newcommand{\fullname}{\textbf{A}daptive \textbf{A}symmetric \textbf{L}abel-guided \textbf{H}ashing}

\usepackage[english]{babel} 

\usepackage[hmarginratio=1:1,left=18mm,right=18mm,top=32mm,columnsep=20pt]{geometry} 
\usepackage[hang, small,labelfont=bf,up,textfont=it,up]{caption} 
\usepackage{booktabs} 

\usepackage{lettrine} 

\usepackage{enumitem} 
\setlist[itemize]{noitemsep} 

\usepackage{abstract} 

\usepackage{titlesec} 
\renewcommand\thesection{\Roman{section}} 
\renewcommand\thesubsection{\roman{subsection}} 
\titleformat{\section}[block]{\large\scshape\centering}{\thesection.}{1em}{} 
\titleformat{\subsection}[block]{\large}{\thesubsection.}{1em}{} 

\usepackage{fancyhdr} 
\usepackage{titling} 

\usepackage{hyperref} 

\pretitle{\begin{center}\Huge\bfseries} 
\posttitle{\end{center}} 
\title{Adaptive Asymmetric Label-guided Hashing for Multimedia Search} 
\author{%
\textsc{Yitian Long} \\[1ex] 
\normalsize Macau University of Science and Technology, Macau, China\\
\normalsize \href{mailto:schuylerlong@163.com}{schuylerlong@163.com} 
}
\date{\today} 
\renewcommand{\maketitlehookd}{%
\begin{abstract}
With the rapid growth of multimodal media data on the Web in recent years, hash learning methods as a way to achieve efficient and flexible cross-modal retrieval of massive multimedia data have received a lot of attention from the current Web resource retrieval research community. Existing supervised hashing methods simply transform label information into pairwise similarity information to guide hash learning, leading to a potential risk of semantic error in the face of multi-label data. In addition, most existing hash optimization methods solve NP-hard optimization problems by employing approximate approximation strategies based on relaxation strategies, leading to a large quantization error. In order to address above obstacles, we present a simple yet efficient \fullname, named \ourmodel, for Multimedia Search. Specifically, \ourmodel~is a two-step hashing method. In the first step, we design an association representation model between the different modality representations and semantic label representation separately, and use the semantic label representation  as an intermediate bridge to solve the semantic gap existing between different modalities. In addition, we present an efficient discrete optimization algorithm for solving the quantization error problem caused by relaxation-based optimization algorithms. In the second step, we leverage the generated hash codes to learn the hash mapping functions. The experimental results show that our proposed method achieves optimal performance on all compared baseline methods. 
\end{abstract}
}

\begin{document}
\maketitle


\section{Introduction}
Since entering the era of big data, mobile Internet and IoT platforms generate huge amount of multimodal data, such as video, audio, text, images, etc., every day. However, it is very difficult to retrieve the information of interest to users quickly from it. One is due to the huge scale of data, large data dimension and complex data structure, which leads to the inability of effective storage and analysis; second, the problem of ``heterogeneous gaps'' between different modalities due to the wide variety of data types and modalities, and the ``semantic gaps'' between low-order features and high-order semantics, which makes it difficult to correlate them. Therefore, how to establish effective large-scale, multi-modal, multi-granularity, multi-scale data efficient indexing mechanism to achieve efficient storage and search of massive and complex data has become a highly hot issue in industry and academia. 

Nearest Neighbor Search (NNS) is a classical data search technique in the field of information retrieval, which aims to accurately retrieve the nearest/most similar sample to the query sample in the database and is widely used in many fields such as data mining, video analysis, and information retrieval. However, nearest neighbor search retrieval is difficult to be accepted in massive data due to its high time consumption. Therefore, Approximate Nearest Neighbor Search (ANNS) meets the retrieval needs of massive data in practical applications by finding potentially similar and not-exactly-similar samples and significantly improving the retrieval efficiency by sacrificing partial precision. It can effectively replace the nearest neighbor search technique by balancing the retrieval accuracy and efficiency. Hashing, as a branch of ANNS, uses hardware-friendly exclusive OR (XOR) instead of traditional matrix computation for efficient Hamming distance metrics during retrieval by mapping high-dimensional data features into a low-dimensional binary space with guaranteed consistent data similarity through a hash function.

However, in the large-scale multimodal data environment, the single-modal data retrieval capability can no longer meet the demand for information retrieval functions, and cross-modal retrieval between different modalities is needed. Although the feature expressions and presentations may be completely different between different modalities, there exists semantic consistency, because when a person views a photo, it is easy to correspond the text that is consistent with the photo description to it and vice versa. Cross-modal retrieval means that given a retrieval sample of one modality, another modality data sample with semantic similarity is returned, such as text search for image, image search for text, etc. Therefore, hash learning methods for large-scale multimodal data retrieval have become a new hot research problem, whose core idea is to map different modalities into a unified Hamming space, solve the ``heterogeneous gap'' problem between different modalities, and achieve efficient cross-modal retrieval.

The evolution of hash learning can be broadly outlined in two ways: data-independent hashing and data-dependent hashing. \textbf{Data-independent hashing}, as the name implies, mapping hash (hash function) is independent of the data and is obtained independently of the data. In other words, the generation process of hash functions is based on manual design or random mapping approach. Representative works are Locality-Sensitive Hashing (LSH) and its variant models. Since the generation of the hash function is not directly related to the data itself, it requires the participation of experts in the relevant fields to generate better quality hash codes, for example, to do medical image data retrieval requires the participation of relevant imaging physicians, because it requires professional physicians to guide the processing rules of the hash function on the relevant domain data, and requires longer hash code bits to obtain more satisfactory retrieval results (usually longer than 1024 bits). This type of hash learning method cannot be widely used in practical applications because of the limitation of the professional field and the long hash code bits. \textbf{Data-dependent hashing}, which is the opposite of data-independent hash learning, has gradually become mainstream with the development of the field of data mining and pattern recognition based on data-dependent hashing techniques. Learning potential knowledge from data improves the discriminative power of the hash function and enhances the semantics of the generated hash code, which makes it possible to obtain high retrieval performance even for short-bit hash codes (usually shorter than 128 bits) and has received wide attention from academia and industry. 

Data-dependent hashing requires learning the relevant mapping functions (hash functions) through the potential knowledge and feature distributions embedded in the data, so it is also known as learning to hashing (L2H), which can be roughly classified into two categories: unsupervised and supervised hashing. Unsupervised hashing usually generates hash codes by mining the data itself for potential correlations or inter-modal feature similarities without label supervision. In contrast, supervised hash learning generically possesses better retrieval performance than unsupervised hashing due to the involvement of semantic labels. Most existing supervised hash learning methods are based on a one-step learning strategy, in other words, hash codes and hash functions are learned and optimized through a unified objective function, which inevitably has several drawbacks: (1) increasing the complexity of optimizing and solving the objective function, resulting in the inability to obtain the optimal solution; (2) the inability to flexibly change the hash mapping function, because using a new hash mapping function will result in retraining the entire objective function. In addition, most hash learning methods leverage the relaxation-based optimization strategies to solve the NP-hard problem posed by discrete hash codes, leading to a large quantization error that affect the retrieval performance of the generated hash codes. In addition to the above limitations, there are still a important issue that need to be addressed, i.e., The semantic gap between differentmodalities is not well bridged. 

In order to address the above limitations, we present a two-step \fullname~(\ourmodel). Specifically, in the first step, we construct a multi-label common semantic space and use this space as a bridge to bridge the semantic gap that exists between different modalities. At the same time, in order to explore the complex nonlinear relationships within different modalities, we use kernelization operations to improve the characterization ability of different modalities. In addition, in the optimization phase, we use discrete optimization strategies to solve the quantization error problem caused by the relaxation strategy. As a result, the semantic and discriminative power of hash codes will be substantially enhanced. In the second step, we use the hash codes generated in the first stage to guide the learning of the hash mapping function.

The main contributions of the paper can be summarized as follows:
\begin{enumerate}
\item An adaptive asymmetric hash learning framework with a common multi-semantic space as a bridge is proposed to establish the association between different modalities and the semantic space, so as to effectively bridge the heterogeneous gap existing between different modalities and improve the discriminative power of the generated hash codes.
\item A discrete optimization strategy based on the augmented Lagrangian method is proposed to solve the quantization error caused by the relaxation strategy and improve the discriminative power of the generated hash codes.
\item The experimental results show that our proposed method achieves optimal performance on two publicly available datasets.
\end{enumerate}

The paper is organized as follows, in Section~\ref{sec:2}, we briefly review the work closely related to this paper. Section~\ref{sec:3} describes our model in detail. Section~\ref{sec:4} gives the results of the experiments, and Section~\ref{sec:5} concludes the work.

\section{Related Works}~\label{sec:2}
The existing hash learning functions can be broadly classified according to the learning paradigm into unimodal hash and cross-modal hash~\cite{DBLP:journals/pami/WangZSSS18}. Unimodal hashing considers only a single modality (e.g., image, text, audio, or video) in the retrieval query and model training phases by maintaining similarity information in the Hamming space of a single modality, which often applied to visual search or text search tasks. In cross-modal hash retrieval, the query is a single modality and the training process consists of multi-modal modalities, which is mainly applied to large-scale media data retrieval, such as text retrieval image or image retrieval text tasks.

\subsection{Unimodal Hashing}
Early works on hash learning focused on unimodal retrieval field~\cite{DBLP:conf/cvpr/ShenSLS15,DBLP:journals/tcyb/LiuDDLL16,DBLP:journals/access/YuanDH17,DBLP:conf/ijcai/LiLDLG18,DBLP:conf/cvpr/ZhuangLSR16,DBLP:conf/aaai/ZhangXLLH19,DBLP:journals/tmm/ZhangZLCW20,DBLP:journals/tmm/MaGLHLY20}. For example, SH~\cite{DBLP:journals/pami/HeoLHCY15} proposes a new hypersphere based hash function that maps more spatially coherent data points into binary codes. Representative works include FSDH~\cite{DBLP:journals/pami/GuiLSTT18} embeds the semantic information of different classes of tags on the corresponding hash codes by regression strategy. CSQ~\cite{DBLP:journals/pami/ZhouYWLLT16} accommodates new image representations by constructing a flexible indexing structure independent of any image descriptor training set. OEH~\cite{DBLP:conf/aaai/LiuJWL16} computes the given order relationship embedded between data points in order to learn the binary code that preserves the ranking. SDAH~\cite{DBLP:journals/ijon/YangRHLZL20} proposes an asymmetric learning framework with different dimensions to generate high-quality image hash codes by considering the information capacity problem of image representation and deep label embedding. FISH~\cite{DBLP:journals/tip/ChenLWGX22} consists of two modules, where the spatial filtering module is responsible for solving the fine-grained feature extraction and the feature filtering module is responsible for solving the feature refinement problem. SCRATCH~\cite{DBLP:journals/tcsv/ChenLLNZX20} finds a potential semantic space using collective matrix decomposition of kernelized features and semantic embedding labels to maintain intra- and inter-modal similarity. DLTH~\cite{DBLP:journals/tip/LiangPLLY22} captures the relative similarity of the new triplets by introducing more triplets and a new listed triplet loss. CUDH~\cite{DBLP:journals/ijon/GuWZYY019} recursively learns discriminative clustering through a soft clustering model and generates binary codes with high similarity responses. EDDH~\cite{DBLP:journals/ipm/YangYHSL21} generates robust hash codes by considering the visual relationships between image pairs while combining semantic labeling information to generate a fine-grained visual-semantic pair similarity matrix.

\subsection{Cross-modal Hashing}
Cross-modal hash learning is mainly to achieve efficient retrieval between different modalities, such as text retrieve images or image retrieves text. Cross-modal hashing can be broadly classified according to the learning paradigm: unsupervised cross-modal hashing and supervised cross-modal hashing.

\subsubsection{Unsupervised Cross-modal Hashing}
Unsupervised hashing accomplishes the generation of hash codes by mining inter- and intra-modal correlations of different modalities without the assistance of label semantics~\cite{DBLP:conf/icmi/HuangMJ19,DBLP:journals/ijon/WangZCLG20,DBLP:journals/cin/LiLT0MY21,DBLP:journals/cogcom/YuWZ22,DBLP:conf/icassp/MikriukovRD22}. Representative works include MGCMH~\cite{DBLP:journals/mta/XieZC16} integrates multigraph learning and hash function learning into a joint framework using an unsupervised learning paradigm to uniformly map data from different modalities into the same hash space. UDCMH~\cite{DBLP:conf/ijcai/WuLHLDZS18} is a model based on deep learning and matrix decomposition with binary latent factor fusion for multimodal data hash code generation in a self-learning manner. DJSRH~\cite{DBLP:conf/iccv/SuZZ19} learns the original neighborhood information of different modalities by constructing a joint semantic affinity matrix, while the proposed reconstruction framework maximally reconstructs the above joint semantic relations to generate robust hash codes. UCH~\cite{DBLP:conf/aaai/LiDWXL19} proposes a novel unsupervised biorthogonal network in which the outer-loop network learns uniform representations of different modalities and the inner-loop network generates hash codes. JDSH~\cite{DBLP:conf/sigir/LiuQGZY20} proposes a novel unsupervised cross-modal joint learning model by constructing joint modal similarity matrices to fully preserve cross-modal semantic associations between instances and distribution-based similarity decision and weighting (DSDW) sampling and weighting schemes in order to generate robust hash codes. DCSH~\cite{DBLP:journals/tip/HoangDNC20} is a two-step unsupervised cross-modal hash learning method that decouples the optimization into two parts: binary optimization and hash function learning. CLIP4Hashing~\cite{DBLP:conf/mir/ZhuoLHHL22} employs a CLIP model to construct a hash learning framework, which is utilized to solve the problem of heterogeneous divide between different modalities in Hamming space.

\subsubsection{Supervised Cross-modal Hashing}
Compared to unsupervised cross-modal hashing, supervised cross-modal hashing using label information can significantly increase the semantics of hash codes and thus improve the retrieval performance of hash codes~\cite{DBLP:journals/tmm/DingFHXP17,DBLP:journals/tip/MandalCB19,DBLP:journals/pr/WangWHGT20,DBLP:journals/pr/LiongLT18,DBLP:journals/tip/JinLHQT18,DBLP:journals/ijon/LinS22,DBLP:journals/corr/abs-2202-10232}. Representative works include LBMCH~\cite{DBLP:conf/sigir/WangLWZZ15} describes the semantic correspondence across schemas by bridging the gaps that exist in different hash spaces. RCMH~\cite{DBLP:conf/sigir/MoranL15} is a three-step hash learning model in which, in one step, each training image is assigned a hash code based on the hyperplane learned in the previous iteration; in the second step, the binary bits are smoothed by a graph-regularized representation so that similar data points have similar bits; and in the third step, a set of binary classifiers are trained to predict the regularized bits with maximum residual. SMFH~\cite{DBLP:conf/ijcai/LiuJWH16} is a supervised cross-modal hash learning method, which solves the multimodal hash learning problem by decomposing the collective non-negative matrix of different modalities. CMFH~\cite{DBLP:journals/tip/TangWS16} is a cross-modal hash learning method based on collective matrix decomposition, which considers both the label consistency of different modalities and the local geometric consistency of each modality. NRDH~\cite{DBLP:conf/sigir/YangL0H20} reconstructs the raw information of different modes by a deep nonlinear descriptor of the common latent representations in reverse, while combining it with a hash learning framework to generate robust hash codes. DDASH~\cite{DBLP:journals/kbs/QiangWLXM20} is an asymmetric hash learning method, which learns the hash codes of database instances by matrix decomposition strategy and the hash codes of query instances by using deep hash function. This not only can take full advantage of the information of large-scale data, but also can reduce the training time of the model. NSDH~\cite{DBLP:journals/kbs/YangYRZHLL21} proposes an asymmetric nonlinear hash learning framework that can incrementally mine the deep semantic information of different modalities layer by layer to generate high-quality hash codes. CMCH~\cite{DBLP:journals/ipm/AnLZZL22} generates discriminative hash codes by progressively mining the structural consensus and informative transformation semantics of different modalities. SASH~\cite{DBLP:journals/tip/ShiNLZY22} mines the relevance of semantic labels by maintaining the consistency of feature and label spaces and uses this relevance to optimize the similarity matrix. Although the above cross-modal supervised methods achieve good retrieval performance, they neglect to fully exploit the potential uniform geometric structure representation of different modalities and the consideration of semantic information completeness issues, resulting in degraded retrieval performance. Therefore, in this paper, we improve the retrieval performance of hash codes by mining the potential uniform integrity representation information of different modalities.


\begin{table*}
\centering
  \caption{Notations.}
  \label{tab:Notations}
  \begin{tabular}{c|c}
    \toprule
    Notation			 & Explanations\\
    \midrule
    $\mathbf{X}^{(m)}\in\mathbb{R}^{d_m\times n}$   & The $m$-th modality representation\cr
    $\phi(\mathbf{X}^{(m)})\in\mathbb{R}^{q\times n}$   & The $m$-th modality kernelized representation\cr
    $\mathbf{V}^{(m)}\in\mathbb{R}^{q\times q}$     & The $m$-th modality-specific space\cr
    $\mathbf{U}\in\mathbb{R}^{q\times n}$           & The common space\cr
    $\mathbf{B}\in\mathbb{R}^{k\times n}$   		& The learned hash codes\cr
    $\mathbf{R}\in\mathbb{R}^{k\times c}$           & Linear projection\cr
    $\mathbf{S}\in\mathbb{R}^{n\times n}$           & Similarity matrix\cr
    $\mathbf{C}\in\mathbb{R}^{k\times q}$           & Linear mapping matrix\cr
    $\mathbf{L}\in\mathbb{R}^{c\times n}$           & Semantic labels\cr
    $\mathbf{W}^{(m)}\in\mathbb{R}^{k\times q}$     & The $m$-th modality hash function\cr
    $n$    				 & Number of instances\cr
    $k$                  & Length of hash codes\cr
    $c$                  & Number of categories\cr
    \bottomrule
  \end{tabular}
\end{table*}

\section{Our proposed \ourmodel}~\label{sec:3}
\subsection{Notation}
Let $\mathbb{I}=\{i_i\}_{i=1}^n$ be the dataset with $n$ instances, and $i_i=\{\mathbf{X}_i^{(m)}\}_{m=1}^M$, $m$ is the $m$-th modality. For convenience, in this paper we will only consider data from two modalities, i.e., visual modality and text modality. $\mathbf{L}\in\mathbb{R}^{c\times n}$ is the semantic information, $c$ is the number of categories. In this paper, we denote $\mathbf{X}^{1}$ and $\mathbf{X}^{2}$ are visual modality and text modality, respectively. $\mathbf{B}\in\mathbb{R}^{k\times n}$ is the hash codes, $k$ is the length of hash codes. The notations used in \ourmodel~are listed in Table.~\ref{tab:Notations}. 

\subsection{The first step of \ourmodel}
\subsubsection{Kernelization}~\label{sec:Kernel}
The kernelization technique is one of the techniques often used in the field of pattern recognition to mine the non-linear complex key within the data. Therefore, to better capture the potential nonlinear relationships between different modalities, the paper uses an expression based on RBF kernel mapping. Specifically, the RBF kernel mapping-based expression makes the kernelized features more discriminative by mining the nonlinear relationships among data, and for a sample $x_i$, its kernelized feature $\phi(x_i)$ can be expressed as,
\begin{equation}\label{eq:Kernel}
\phi(x_i)=\left[\exp(\frac{-||x_i-a_1||_2^2}{2\rho^2}),...,\exp(\frac{-||x_i-a_q||_2^2}{2\rho^2})\right]^\top,
\end{equation}
where $a_q$ represents the randomly selected $q$ anchor instances and $\rho$ is the width.

\subsubsection{Modality-Specific Association Learning}
The different modalities describe the same common space from different perspectives, then these modalities should exist in the common space $\mathbf{U}\in\mathbb{R}^{q\times n}$. In addition, the different modalities also have their specific feature space, i.e. modality-specific space ${\mathbf{V}_m}\in\mathbb{R}^{q\times q}$. Then, we obtain,
\begin{equation}\label{eq:MSA}
\begin{aligned}
&\underset{\mathbf{U,V}_m}{\min}~\mu_m||\phi(\mathbf{X}^{(m)})-{\mathbf{V}_m}\mathbf{U}||_F^2\\
&s.t.~{\mathbf{V}_m}^\top\mathbf{V}_m=\mathbf{I}_q, \sum_m \mu_m=1, \mu_m>0,
\end{aligned}
\end{equation}
where $\mu_m$ is a weight parameter for the $m$-th modality. In the most existing hashing methods, the weight parameter $\mu_m$ is fixed. However, the degree of contribution of different modalities to the common latent representation is different, and manual adjustment of the parameters firstly cannot accurately determine the degree of contribution of different modalities, i.e., the optimal values of $\mu_m$. Secondly, searching for the optimal parameters leads to huge time consumption and cannot be applied in practical scenarios. Therefore, in this paper, we propose an \textbf{adaptive} learning scheme to obtain the values of $\mu_m$. In order to avoid ``winner-take-all'' phenomenon, we introduce a exponential parameter $\zeta$ to smooth the contribution of different modalities, we obtain,
\begin{equation}\label{eq:MSA_2}
\begin{aligned}
&\underset{\mu_m,\mathbf{U,V}_m}{\min}~{\mu_m}^{\zeta}||\phi(\mathbf{X}^{(m)})-{\mathbf{V}_m}\mathbf{U}||_F^2\\
&s.t.~{\mathbf{V}_m}^\top\mathbf{V}_m=\mathbf{I}_q, \sum_m \mu_m=1, \mu_m>0,
\end{aligned}
\end{equation}
where $\zeta\geq 2$. In this way, we can learn the optimal weighting parameters. In addition, we associate the common space $\bf U$ with the hash space $\bf B$ by the use of the linear mapping $\mathbf{C}\in\mathbb{R}^{k\times q}$. The formula is,
\begin{equation}~\label{eq:UCB}
\begin{aligned}
&\underset{\mathbf{U,C,B}}{\min}~||\mathbf{B}-\mathbf{CU}||_F^2.
\end{aligned}
\end{equation}

\subsubsection{Multi-Semantic Space Learning}
The core problem of cross-modal retrieval is to solve the problem of heterogeneous gap existing between different modalities. We address this problem from another perspective, i.e., (i) constructing a reasonable common space; (ii) then associating the different modalities with it. This strategy has the following advantages: (1) it bridges the heterogeneity gap among different modalities; (2) it can be easily extended to multiple modalities. Moreover, the problem of semantically accurate representation of multi-label needs to be addressed, because constructing the multi-label pair similarity matrix directly leads to semantic errors. For example, there are three samples $x_1,x_2,x_3$, corresponding to labels $y_1=[0,1,1,1,1,1,1]$,$y_2=[0,1,1,1,1,1,1]$,$y_3=[1,1,0,0,0,0,0]$, and the reconstructed pair-wise similarity information is $S_{12}=S_{23}=S_{13}=1$. However, the distance between $x_1$ and $x_2$ is smaller than the distance between $x_1$ and $x_3$. In the above development, we leverage a linear projection $\mathbf{R}\in\mathbb{R}^{k\times c}$ to embed the multi-label vectors to a final common space, i.e., the Hamming space $\bf B$. In general, associating label semantic information to each category increases the semantics of the learned hash codes,
\begin{equation}\label{eq:LabelC}
\begin{aligned}
&\underset{\mathbf{B,R}}{\min}~||\mathbf{RL}-\mathbf{B}||_F^2\\
&~s.t.~\mathbf{B}\in\{-1,1\}^{k\times n}.
\end{aligned}
\end{equation}

\subsubsection{Hash Code Learning}
The KSH Method is a very well-known hash learning method that proposes a semantic preservation principle based on hash code learning, as
\begin{equation}~\label{eq:KSH}
\begin{aligned}
&\underset{\mathbf{B}}{\min}~||\mathbf{B}^\top\mathbf{B}-k\mathbf{S}||_F^2\\
&~s.t.~\mathbf{B}\in\{-1,1\}^{k\times n}.
\end{aligned}
\end{equation}

However, Equation~\eqref{eq:KSH} has several drawbacks: (i) how to construct the $n\times n$ pairwise similarity matrix $\bf S$ efficiently; (ii) how to solve the discrete optimization efficiently. In response to the drawback (i), we replace $\bf S$ with $2\mathbf{L}^\top\mathbf{L}-\mathbf{11}^\top$, where $\mathbf{1}$ means a vector with all elements being one, then the time cost $\mathcal{O}(n^2)$ can be reduced to $\mathcal{O}(n)$. For the drawback (ii), we construct learning-efficient and efficient asymmetric hash learning architectures, and a large number of researchers have shown that the use of asymmetric hash learning frameworks significantly outperforms symmetric hash learning frameworks in terms of both learning efficiency and retrieval accuracy. Combining the Eq.~\eqref{eq:UCB}, we substitute one of $\bf B$ in Eq.~\eqref{eq:UCB} with multi-modal matrix $\mathbf{CU}$ to construct an asymmetric learning framework, as
\begin{equation}~\label{eq:ASYH}
\begin{aligned}
&\underset{\mathbf{C,B,U}}{\min}~||\mathbf{B}^\top\mathbf{CU}-k\mathbf{S}||_F^2+\alpha||\mathbf{B}-\mathbf{CU}||_F^2\\
&~s.t.~\mathbf{B}\in\{-1,1\}^{k\times n},
\end{aligned}
\end{equation}
where $\alpha$ is a trade-off parameter. In order to enhance the semantic information of the learned hash codes, the second $\bf B$ in Eq.~\eqref{eq:UCB} also needs to be replaced with Eq.~\eqref{eq:LabelC}, then we obtain,
\begin{equation}~\label{eq:ASYH}
\begin{aligned}
&\underset{\mathbf{R,C,B,U}}{\min}~||({\mathbf{RL}})^\top\mathbf{CU}-k\mathbf{S}||_F^2+\alpha||\mathbf{B}-\mathbf{CU}||_F^2+\beta||\mathbf{RL}-\mathbf{B}||_F^2\\
&~s.t.~\mathbf{B}\in\{-1,1\}^{k\times n},
\end{aligned}
\end{equation}
where $\beta$ is a balance parameter.

\subsection{Optimization}
Combing Eq.~\eqref{eq:ASYH} and Eq.~\eqref{eq:MSA}, we obtain the overall objective function,
\begin{equation}\label{eq:Over}
\begin{aligned}
&\underset{\mu_m,\mathbf{R,C,B,U,V}_m}{\min}~||({\mathbf{RL}})^\top\mathbf{CU}-k\mathbf{S}||_F^2+\alpha||\mathbf{B}-\mathbf{CU}||_F^2\\
&+\beta||\mathbf{RL}-\mathbf{B}||_F^2+{\mu_m}^{\zeta}||\phi(\mathbf{X}^{(m)})-{\mathbf{V}_m}\mathbf{U}||_F^2+\eta\mathcal{R}(\mathbf{RL,U})\\
&~s.t.~\mathbf{B}\in\{-1,1\}^{k\times n},~{\mathbf{V}_m}^\top\mathbf{V}_m=\mathbf{I}_q, \sum_m \mu_m=1, \mu_m>0,
\end{aligned}
\end{equation}
where $\eta$ is a regularization parameter, the term $\mathcal{R}(\mathbf{R,U})=||\mathbf{RL}||_F^2+||\mathbf{U}||_F^2$ is a regularization term that avoids overfitting.

Solving Eq.~\eqref{eq:Over} is actually an NP-Hard problem due to the discrete value optimization. Many methods are optimized by a two-step approach, i.e., first making an approximation to the discrete values and then solving for the approximation; In the second step, the approximation variables are processed using the \textit{sgn} function. Such an approach solves the NP-hard problem brought by the discretized value solution, but it also causes a large amount of quantization error, which affects the retrieval performance. Some methods use the DDC solving strategy to solve hash codes bit-by-bit. Although this method solves the quantization loss brought by the relaxation strategy, it takes $k$ iterations to optimize a hash code of length $k$ bits, which leads to a large amount of time consumption.

In this paper, we use a discrete optimization strategy to generate discrete hash codes directly in one step to solve the problems of quantization error and excessive time consumption. Specifically, we fix the other variables by solving for one of them. The overall optimization process of Eq.~\eqref{eq:Over} is as follows,

\textbf{$\mu_m$-step}: We fix the the other variables, i.e., $\mathbf{R,C,B,U,V}_m$, the updating for variable $\mu$ can be reformulated as,
\begin{equation}\label{eq:mu1}
\begin{aligned}
&\underset{\mu_m}{\min}~{\mu_m}^{\zeta}||\phi(\mathbf{X}^{(m)})-{\mathbf{V}_m}\mathbf{U}||_F^2
&\sum_m \mu_m=1, \mu_m>0.
\end{aligned}
\end{equation}

We leverage the Lagrangian multiplier algorithm to reduce the difficulty of solving the variable $\mu_m$, thus the Eq.~\eqref{eq:mu1} can be reformulated as,
\begin{equation}\label{eq:mu2}
\begin{aligned}
&\underset{\mu_m}{\min}~{\mu_m}^{\zeta}||\phi(\mathbf{X}^{(m)})-{\mathbf{V}_m}\mathbf{U}||_F^2-\Xi(\mathbf{1}^\top\mu - 1),
\end{aligned}
\end{equation}
where $\mu=[\mu_1,\mu_2,...,\mu_M]^\top \in\mathbb{R}^{M}$ is the vector of weights for the different modalities.

Setting the derivative with respect to $\Xi$ and $\mu$ to zero, the solution of the variable $\mu_m$ can be written as,
\begin{equation}\label{mu_s}
\begin{aligned}
\mu_m=\frac{\Delta_m^{1/1-\zeta}}{\sum_{m=1}^M\Delta_m^{1/1-\zeta}},
\end{aligned}
\end{equation}
where $\Delta_m=||\phi(\mathbf{X}^{(m)})-{\mathbf{V}_m}\mathbf{U}||_F^2$.

\textbf{$\bf R$-step}: We fix the the other variables, i.e., $\mu_m,\mathbf{C,B,U,V}_m$, the updating for variable $\bf R$ can be reformulated as
\begin{equation}\label{eq:R1}
\begin{aligned}
&\underset{\mathbf{R}}{\min}~||({\mathbf{RL}})^\top\mathbf{CU}-k\mathbf{S}||_F^2+\beta||\mathbf{RL}-\mathbf{B}||_F^2+\eta\mathcal{R}(\mathbf{RL}).
\end{aligned}
\end{equation}

Then, Eq.~\eqref{eq:R1} can be rewritten as,
\begin{equation}\label{eq:R2}
\begin{aligned}
&\underset{\mathbf{R}}{\min}~tr(\mathbf{L}^\top\mathbf{R}^\top\mathbf{CUU}^\top\mathbf{C}^\top\mathbf{RL}-2k\mathbf{L}^\top\mathbf{R}^\top\mathbf{CUS}^\top\\
&+\beta\mathbf{RLL}^\top\mathbf{R}^\top-2\beta\mathbf{BL}^\top\mathbf{R}^\top+\eta\mathbf{RLL}^\top\mathbf{R}^\top).
\end{aligned}
\end{equation}

We compute the deviation w.r.t $\bf R$ to 0. Then, we obtain the closed-form solution as, 
\begin{equation}\label{eq:R_v}
\mathbf{R}=(\mathbf{CUU}^\top\mathbf{C}^\top+(\beta+\eta)\mathbf{I})^{-1}(k\mathbf{CUS}^\top\mathbf{L}^\top+\beta\mathbf{BL}^\top)(\mathbf{LL}^\top)^{-1}
\end{equation}

Then, the term $k\mathbf{CUS}^\top\mathbf{L}^\top$ can be obtained by $2(\mathbf{CUL}^\top){(\mathbf{LL}^\top)}^\top-(\mathbf{CU1})(\mathbf{L1})^\top$. which consumes $\mathcal{O}(n)$.

\textbf{$\bf C$-step}: We fix the the other variables, i.e., $\mu_m,\mathbf{R,B,U,V}_m$, the updating for variable $\bf C$ can be reformulated as,
\begin{equation}\label{eq:C1}
\begin{aligned}
&\underset{\mathbf{C}}{\min}~||({\mathbf{RL}})^\top\mathbf{CU}-k\mathbf{S}||_F^2+\alpha||\mathbf{B}-\mathbf{CU}||_F^2.
\end{aligned}
\end{equation}

Then, Eq.~\eqref{eq:C1} can be rewritten as,
\begin{equation}\label{eq:C2}
\begin{aligned}
&\underset{\mathbf{C}}{\min}~tr(\mathbf{L}^\top\mathbf{R}^\top\mathbf{CUU}^\top\mathbf{C}^\top\mathbf{RL}-2k\mathbf{S}\mathbf{U}^\top\mathbf{C}^\top\mathbf{RL}\\
&-2\alpha\mathbf{BU}^\top\mathbf{C}^\top+\alpha\mathbf{CUU}^\top\mathbf{C}^\top).
\end{aligned}
\end{equation}

We compute the deviation w.r.t $\bf C$ to 0. Then, we obtain the closed-form solution as, 
\begin{equation}\label{eq:C_v}
\mathbf{C}=(\mathbf{RLL}^\top\mathbf{R}^\top+\alpha\mathbf{I})^{-1}(k\mathbf{RLSU}^\top+\alpha\mathbf{BU}^\top)(\mathbf{UU}^\top)^{-1}
\end{equation}

Similar, the term $\mathbf{RLSU}^\top$ can be rewritten as $2(\mathbf{RLL}^\top)(\mathbf{UL}^\top)^\top-(\mathbf{RL1})(\mathbf{U1})^\top$, which consumes $\mathcal{O}(n)$.

\textbf{$\bf B$-step}: We fix the the other variables, i.e., $\mu_m,\mathbf{R,C,U,V}_m$, the updating for variable $\bf B$ can be reformulated as,
\begin{equation}\label{eq:B1}
\begin{aligned}
&\underset{\mathbf{B}}{\min}~\alpha||\mathbf{B}-\mathbf{CU}||_F^2+\beta||\mathbf{RL}-\mathbf{B}||_F^2\\
&~s.t.~\mathbf{B}\in\{-1,1\}^{k\times n}.
\end{aligned}
\end{equation}

Then, Eq.~\eqref{eq:B1} can be rewritten as,
\begin{equation}\label{eq:B2}
\begin{aligned}
&\underset{\mathbf{B}}{\min}~tr((\alpha+\beta)\mathbf{BB}^\top-2\alpha\mathbf{CUB}^\top-2\beta\mathbf{RLB}^\top).
\end{aligned}
\end{equation}

Note that, the term $\mathbf{BB}^\top=nk$ is a constant. Therefore, Eq.~\eqref{eq:B2} can be reformulated as,
\begin{equation}\label{eq:B2}
\begin{aligned}
&\underset{\mathbf{B}}{\max}~tr((\alpha\mathbf{CU}+\beta\mathbf{RL})\mathbf{B}^\top).
\end{aligned}
\end{equation}

The closed-form solution of the variable $\bf B$ is,
\begin{equation}\label{eq:B_v}
\mathbf{B}=\alpha\mathbf{CU}+\beta\mathbf{RL}.
\end{equation}

\textbf{$\bf U$-step}: We fix the the other variables, i.e., $\mu_m,\mathbf{R,C,B,V}_m$, the updating for variable $\bf U$ can be reformulated as,
\begin{equation}\label{eq:U1}
\begin{aligned}
&\underset{\mathbf{U}}{\min}~||({\mathbf{RL}})^\top\mathbf{CU}-k\mathbf{S}||_F^2+\alpha||\mathbf{B}-\mathbf{CU}||_F^2\\
&+{\mu_m}^{\zeta}||\phi(\mathbf{X}^{(m)})-{\mathbf{V}_m}\mathbf{U}||_F^2+\eta\mathcal{R}(\mathbf{U}),
\end{aligned}
\end{equation}

Then, Eq.~\eqref{eq:U1} can be rewritten as,
\begin{equation}\label{eq:U2}
\begin{aligned}
&\underset{\mathbf{U}}{\min}~tr(\mathbf{L}^\top\mathbf{R}^\top\mathbf{CUU}^\top\mathbf{C}^\top\mathbf{RL}-2k\mathbf{SU}^\top\mathbf{C}^\top\mathbf{RL}\\
&+\alpha\mathbf{CUU}^\top\mathbf{C}^\top-2\alpha\mathbf{BU}^\top\mathbf{C}^\top-2{\mu_m}^{\zeta}\phi(\mathbf{X}^{(m)})\mathbf{U}^\top{\mathbf{V}_m}^\top\\
&+{\mu_m}^{\zeta}\mathbf{V}_m\mathbf{UU}^\top{\mathbf{V}_m}^\top+\eta\mathbf{UU}^\top).
\end{aligned}
\end{equation}

The closed-form solution of the variable $\bf U$ is,
\begin{equation}\label{eq:U_v}
\begin{aligned}
&\mathbf{U}=(\mathbf{C}^\top\mathbf{RLL}^\top\mathbf{R}^\top\mathbf{C}+\alpha\mathbf{C}^\top\mathbf{C}+{\mu_m}^{\zeta}{\mathbf{V}_m}^\top\mathbf{V}_m+\eta\mathbf{I})^{-1}\\
&\cdot(k\mathbf{C}^\top\mathbf{RLS}+\alpha\mathbf{C}^\top\mathbf{B}+{\mu_m}^{\zeta}{\mathbf{V}_m}^\top\phi(\mathbf{X}^{(m)}))
\end{aligned}
\end{equation}

\textbf{$\mathbf{V}_m$-step}: We fix the the other variables, i.e., $\mu_m,\mathbf{R,C,B,U}$, the updating for variable $\mathbf{V}_m$ can be reformulated as,
\begin{equation}\label{eq:V1}
\begin{aligned}
&\underset{\mathbf{V}_m}{\min}~{\mu_m}^{\zeta}||\phi(\mathbf{X}^{(m)})-{\mathbf{V}_m}\mathbf{U}||_F^2\\
&s.t.~{\mathbf{V}_m}^\top\mathbf{V}_m=\mathbf{I}_q.
\end{aligned}
\end{equation}

Then, Eq.~\eqref{eq:V1} can be rewritten as,
\begin{equation}\label{eq:V2}
\begin{aligned}
&\underset{{\mathbf{V}_m}^\top\mathbf{V}_m=\mathbf{I}_q}{\min}~tr(-2{\mu_m}^{\zeta}\phi(\mathbf{X}^{(m)})\mathbf{U}^\top{\mathbf{V}_m}^\top+{\mu_m}^{\zeta}\mathbf{V}_m\mathbf{UU}^\top{\mathbf{V}_m}^\top).
\end{aligned}
\end{equation}

We introduce an auxiliary variable $\mathbf{K}_m$, and set $\mathbf{K}_m=\mathbf{V}_m$. Then, Eq.~\eqref{eq:V2} can be rewritten as,
\begin{equation}\label{eq:V3}
\begin{aligned}
&\underset{{\mathbf{V}_m}^\top\mathbf{V}_m=\mathbf{I}_q}{\min}~tr(-2{\mu_m}^{\zeta}\phi(\mathbf{X}^{(m)})\mathbf{U}^\top{\mathbf{V}_m}^\top+{\mu_m}^{\zeta}\mathbf{K}_m\mathbf{UU}^\top{\mathbf{V}_m}^\top)\\
&+\frac{\lambda}{2}||\mathbf{V}_m-\mathbf{K}_m+\frac{{\mathbf{\Lambda_V}_m}}{\lambda}||_F^2,
\end{aligned}
\end{equation}
where $\mathbf{\Lambda_V}_m$ is the difference between the auxiliary variable and target variable, $\lambda>0$ is the trade-off parameter. Then Eq.~\eqref{eq:V3} can be obtained by,
\begin{equation}\label{eq:V4}
\begin{aligned}
&\underset{{\mathbf{V}_m}^\top\mathbf{V}_m=\mathbf{I}_q}{\min}~tr\left(-2{\mu_m}^{\zeta}\phi(\mathbf{X}^{(m)})\mathbf{U}^\top{\mathbf{V}_m}^\top+{\mu_m}^{\zeta}\mathbf{K}_m\mathbf{UU}^\top{\mathbf{V}_m}^\top\right.\\
&\left.-\lambda{\mathbf{V}_m}^\top(\mathbf{K}_m-\frac{{\mathbf{\Lambda_V}_m}}{\lambda})\right).
\end{aligned}
\end{equation}

Eq.~\eqref{eq:V4} can be transformed as,
\begin{equation}\label{eq:V5}
\begin{aligned}
&\underset{{\mathbf{V}_m}^\top\mathbf{V}_m=\mathbf{I}_q}{\max}~tr\left(\mathbf{O}{\mathbf{V}_m}^\top\right),
\end{aligned}
\end{equation}
where $\mathbf{O}=2{\mu_m}^{\zeta}\phi(\mathbf{X}^{(m)})\mathbf{U}^\top-{\mu_m}^{\zeta}\mathbf{K}_m\mathbf{UU}^\top+\lambda\mathbf{K}_m-{\mathbf{\Lambda_V}_m}$. The optimal $\mathbf{V}_m$ can be solved by $\mathbf{V}_m=\mathcal{P}\mathcal{Q}^\top$, where $\mathcal{P}$ and $\mathcal{Q}$ are composed of left-singular and right-singular of $\mathbf{O}$, respectively.

\textbf{$\mathbf{K}_m$-step}: The objective function of $\mathbf{K}_m$ can be obtained as,
\begin{equation}\label{eq:K_v}
\underset{{\mathbf{K}_m}^\top\mathbf{K}_m=\mathbf{I}_q}{\max}~tr({\mathbf{O}_k{\mathbf{K}_m}^\top}),
\end{equation}
where $\mathbf{O}_k=-{\mu_m}^{\zeta}{\mathbf{V}_m}^\top\mathbf{UU}^\top+\lambda\mathbf{V}_m+{\mathbf{\Lambda_V}_m}$. The optimal solution of $\mathbf{K}_m$ is $\mathcal{P}_k{\mathcal{Q}_k}^\top$, where $\mathcal{P}_k$ and $\mathcal{Q}_k$ are composed of left-singular and right-singular of $\mathbf{O}_k$, respectively.

\textbf{$\mathbf{\Lambda_V}_m$-step}: The update rule of $\mathbf{\Lambda_V}_m$ is,
\begin{equation}\label{eq:Lambda_v}
\mathbf{\Lambda_V}_m=\mathbf{\Lambda_V}_m+\lambda(\mathbf{V}_m-\mathbf{K}_m).
\end{equation}

\subsection{The second step of \ourmodel}
Our proposed method is a two-step hashing, so in the second step, we need to learn the hash map hash by the hash code generated in the first step. Hash functions can be linear mapping functions, deep neural networks, support vector machine, etc. In order to accelerate and simplify the learning process of hash function, linear hash mapping function is used in this paper. Specifically, the learning process of hash function can be done by the following,
\begin{equation}
\underset{\mathbf{W}_m}{\min}~||\mathbf{B}-\mathbf{W}_m\phi(\mathbf{X}_m)||_F^2+\omega||\mathbf{W}_m||_F^2,
\end{equation}
where $\omega$ is a regularization parameter. The optimal solution of the variable $\mathbf{W}_m\in\mathbb{R}^{k\times q}$ is,
\begin{equation}\label{eq:W_v}
\mathbf{W}_m=\mathbf{B}\phi(\mathbf{X})^\top(\phi(\mathbf{X})\phi(\mathbf{X})^\top+\omega\mathbf{I})^{-1}.
\end{equation}

\begin{algorithm}[!tb]
\small
\caption{\bf \fullname~(\ourmodel)}
\label{alg:1}
\SetKwInOut{Input}{Input}\SetKwInOut{Output}{Output}
\Input{The multimedia data features $\mathbf{X}^{(m)}$, parameters $\alpha,\beta,\zeta,\eta,\lambda$, and maximal iteration number $T$.}
\Output{Binary code $\bf B$, Hash function $\mathbf{W}_m$.}
\textbf{Procedure:}\\
\textit{\% The first step of \ourmodel}\\
\textbf{Initialization:} Compute the kernel-based data features via Eq.~\eqref{eq:Kernel}, initialize the variables $\mu_m,\mathbf{R,C,B,U,V}_m,\mathbf{K}_m$ and $i=1$.\\
\While{$i<T$ and not converged}{
$t=1$;\\
Update $\mu_m$ via~\eqref{mu_s};\\
Update $\bf R$ via~\eqref{eq:R_v};\\
Update $\bf C$ via~\eqref{eq:C_v};\\
Update $\bf B$ via~\eqref{eq:B_v};\\
Update $\bf U$ via~\eqref{eq:U_v};\\
Update $\mathbf{V}_m$ via~\eqref{eq:V5};\\
Update $\mathbf{K}_m$ via~\eqref{eq:K_v};\\
$i=i+1$;
}
\textit{\% The second step of \ourmodel}\\
Update $\mathbf{W}_m$ via~\eqref{eq:W_v};\\
\end{algorithm}

\subsection{Time Cost Analysis}
In this section, we analyze the time consumption of the variables in the two stages of the \ourmodel, where the objective function involves a total of seven variables, i.e., $\mathbf{R,C,B,U,V}_m,\mathbf{W}_m$. In the training stage, the time cost of solving $\bf R$ is $\mathcal{O}(kqn+k^2q+k^3+kcn+c^2n+kc^2+c^3+kc+k^2c)$, the time cost of solving $\bf C$ is $\mathcal{O}(kcn+k^2c+k^3+kcq+qcn+q^2n+q^3+k^2q+kq^2+kn+qn+kq)$, the time of solving $\bf B$ is $\mathcal{O}(kqn)$, the time complexity of solving $\bf U$ is  $\mathcal{O}(qkc+qcn+q^2k+q^2n+qkn+q^3)$, the time complexity of solving $\mathbf{V}_m$ is $\mathcal{O}(q^2n+q^3)$.  In the inference stage, the time cost of solving $\bf W$ is $\mathcal{O}(q^2n+q^3+kq^2+kqn)$. Assume $q,k,c,T\ll n$, where $T$ is the number of iterations. Therefore, it can be easily concluded that the overall time complexity of our proposed \ourmodel~is linear in the number of instances, i.e., $n$.

\section{Experimental Results}~\label{sec:4}
\subsection{Dataset Descriptions}
We validate our proposed \ourmodel~with two publicly available datasets, i.e., MIRFlickr and NUS-WIDE datasets.
\begin{enumerate}
\item \textbf{MIRFlickr dataset} consists of 25k instances obtained from Flick.com. Each instance is labelled as some of 24 tags. Each text modality data is characterized by a 1,386-dim Bag-of Words (BoW) feature vector and each image modality data is characterized by a 512-dim GIST feature vector.
\item \textbf{NUS-WIDE dataset} contains 269,648 instances with 81 different tags. We select the top 10 most used tags, then obtain a subset consisting of 186,577 instances. Each text modality data is characterized by a 1,000-dim Bag-of Words (BoW) feature vector and each image modality data is characterized by a 500-dim Bag-of-Visual Words (BoVW) feature vector.
\end{enumerate}

\begin{table}
\setlength{\abovecaptionskip}{10pt}
\small
  \caption{Performance comparison on MIRFlickr-25K measured by mAP.}
  \label{tab:mAP_MIR}
  \begin{tabular}{c|l|ccccc}
    \toprule
    Task &Method & 16 bits& 32 bits&64 bits&128 bits\\
    \midrule
    \multirow{10}{0.7cm}{I$\rightarrow$T}& CMFH    & 0.5687 & 0.5680 & 0.5685 & 0.5687\cr
    								     & SCM-seq & 0.6373 & 0.6478 & 0.6537 & 0.6611\cr
    								     & SePH-km & 0.6685 & 0.6818 & 0.6830 & 0.6873\cr
    								     & FSH     & 0.6016 & 0.6149 & 0.6194 & 0.6242\cr
    								     & TECH    & 0.7304 & 0.7478 & 0.7553 & 0.7615\cr
    								     & SRLCH   & 0.6234 & 0.6328 & 0.6271 & 0.6561\cr
    								     & MTFH    & 0.7489 & 0.7533 & 0.7578 & 0.7631\cr
    								     & LESGH   & 0.7512 & 0.7598 & 0.7621 & 0.7759\cr	
    								     &\ourmodel & \textbf{0.7531} & \textbf{0.7638} & \textbf{0.7717} & \textbf{0.7845}\cr
    \hline
    \multirow{10}{0.7cm}{T$\rightarrow$I} & CMFH    & 0.5615 & 0.5606 & 0.5606 & 0.5608\cr
    								     & SCM-seq & 0.6206 & 0.6298 & 0.6372 & 0.6427\cr
    								     & SePH-km & 0.7076 & 0.7212 & 0.7293 & 0.7348\cr
    								     & FSH     & 0.5979 & 0.6114 & 0.6186 & 0.6251\cr
    								     & TECH    & 0.8000 & 0.8241 & 0.8361 & 0.8449\cr
    								     & SRLCH   & 0.6052 & 0.6161 & 0.6412 & 0.6021\cr
    								     & MTFH    & 0.7933 & 0.8016 & 0.8147 & 0.8264\cr
    								     & LESGH   & 0.8126 & 0.8185 & 0.8203 & 0.8290\cr			     
    								     &\ourmodel & \textbf{0.8133} & \textbf{0.8253} & \textbf{0.8262} & \textbf{0.8328}\cr
    \bottomrule
  \end{tabular}
\end{table}

\begin{table}
\setlength{\abovecaptionskip}{10pt}
\small
  \caption{Performance comparison on NUS-WIDE measured by mAP.}
  \label{tab:mAP_NUS}
  \begin{tabular}{c|l|ccccc}
    \toprule
    Task &Method & 16 bits& 32 bits&64 bits&128 bits\\
    \midrule
    \multirow{10}{0.7cm}{I$\rightarrow$T} & CMFH    & 0.3437 & 0.3399 & 0.3409 & 0.3440\cr
    								     & SCM-seq & 0.5120 & 0.5422 & 0.5488 & 0.5483\cr
    								     & SePH-km & 0.5537 & 0.5627 & 0.5622 & 0.5698\cr
    								     & FSH     & 0.3732 & 0.3894 & 0.4014 & 0.4084\cr
    								     & TECH    & 0.6221 & 0.6375 & 0.6450 & 0.6415\cr
    								     & SRLCH   & 0.3578 & 0.3604 & 0.3671 & 0.3752\cr
    								     & MTFH    & 0.5921 & 0.6087 & 0.6246 & 0.6221\cr
    								     & LESGH   & 0.6221 & 0.6336 & 0.6389 & 0.6451\cr				     
    								    &\ourmodel & \textbf{0.6298} & \textbf{0.6388} & \textbf{0.6497} & \textbf{0.6507}\cr
    \hline
    \multirow{10}{0.7cm}{T$\rightarrow$I} & CMFH     & 0.3498 & 0.3435 & 0.3486 & 0.3529\cr
    								     & SCM-seq  & 0.4836 & 0.5067 & 0.5141 & 0.5161\cr
    								     & SePH-km  & 0.6407 & 0.6515 & 0.6608 & 0.6651\cr
    								     & FSH      & 0.3717 & 0.3835 & 0.3973 & 0.4007\cr
    								     & TECH     & 0.7665 & 0.7858 & 0.7975 & 0.7950\cr
    								     & SRLCH    & 0.3600 & 0.3641 & 0.3514 & 0.3676\cr
    								     & MTFH     & 0.6791 & 0.6973 & 0.7187 & 0.7191\cr
    								     & LESGH    & 0.7474 & 0.7643 & 0.7812 & 0.7884\cr								     
    								    &\ourmodel  & \textbf{0.7698} & \textbf{0.7942} & \textbf{0.8032} & \textbf{0.8057}\cr
    \bottomrule
  \end{tabular}
\end{table}

%

\begin{figure*}
 \centering
  \subfigure{
   \includegraphics[width=3in]{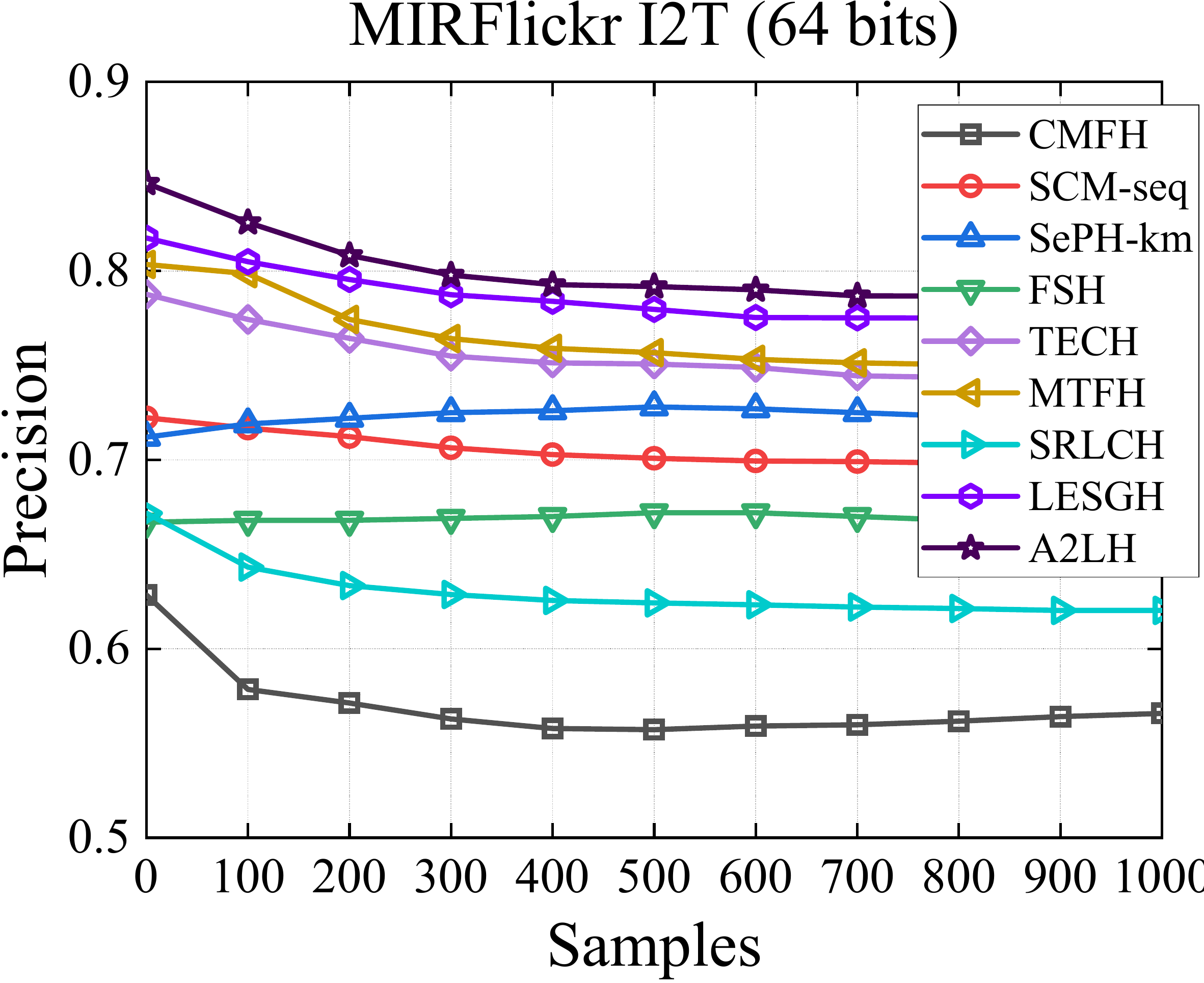}
   }
  \subfigure{
 \includegraphics[width=3in]{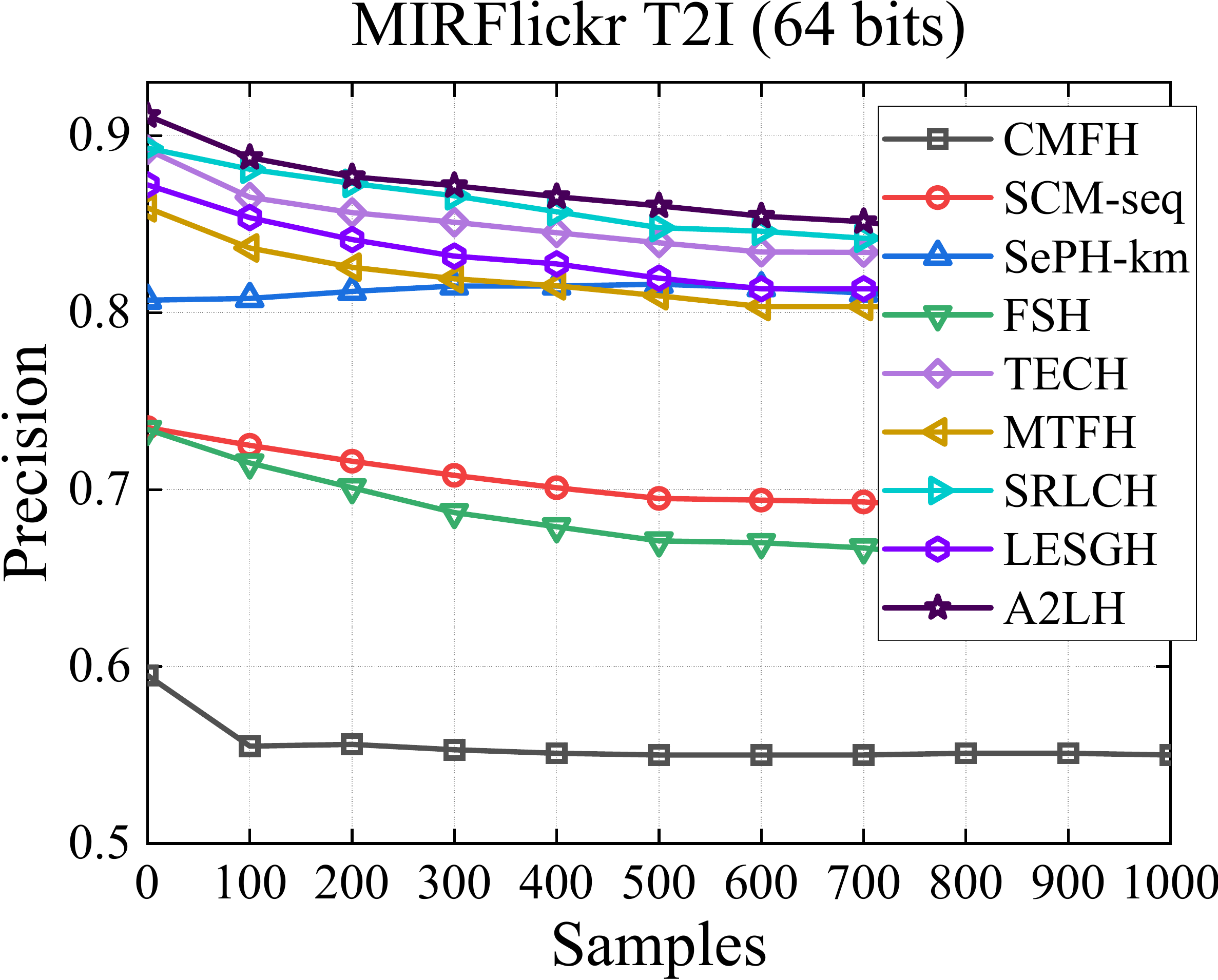}
 }
 
  \subfigure{
   \includegraphics[width=3in]{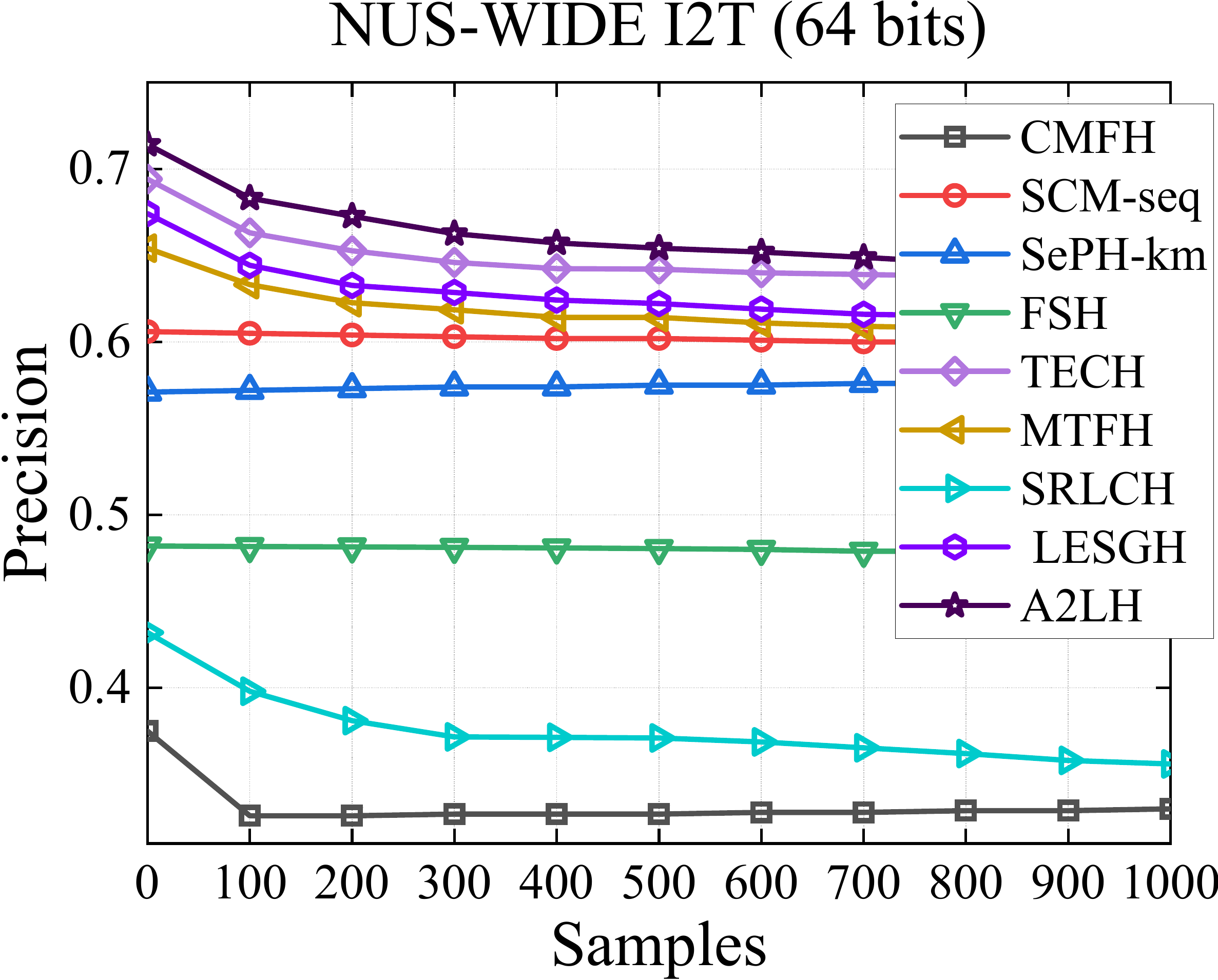}
   }
   \subfigure{
   \includegraphics[width=3in]{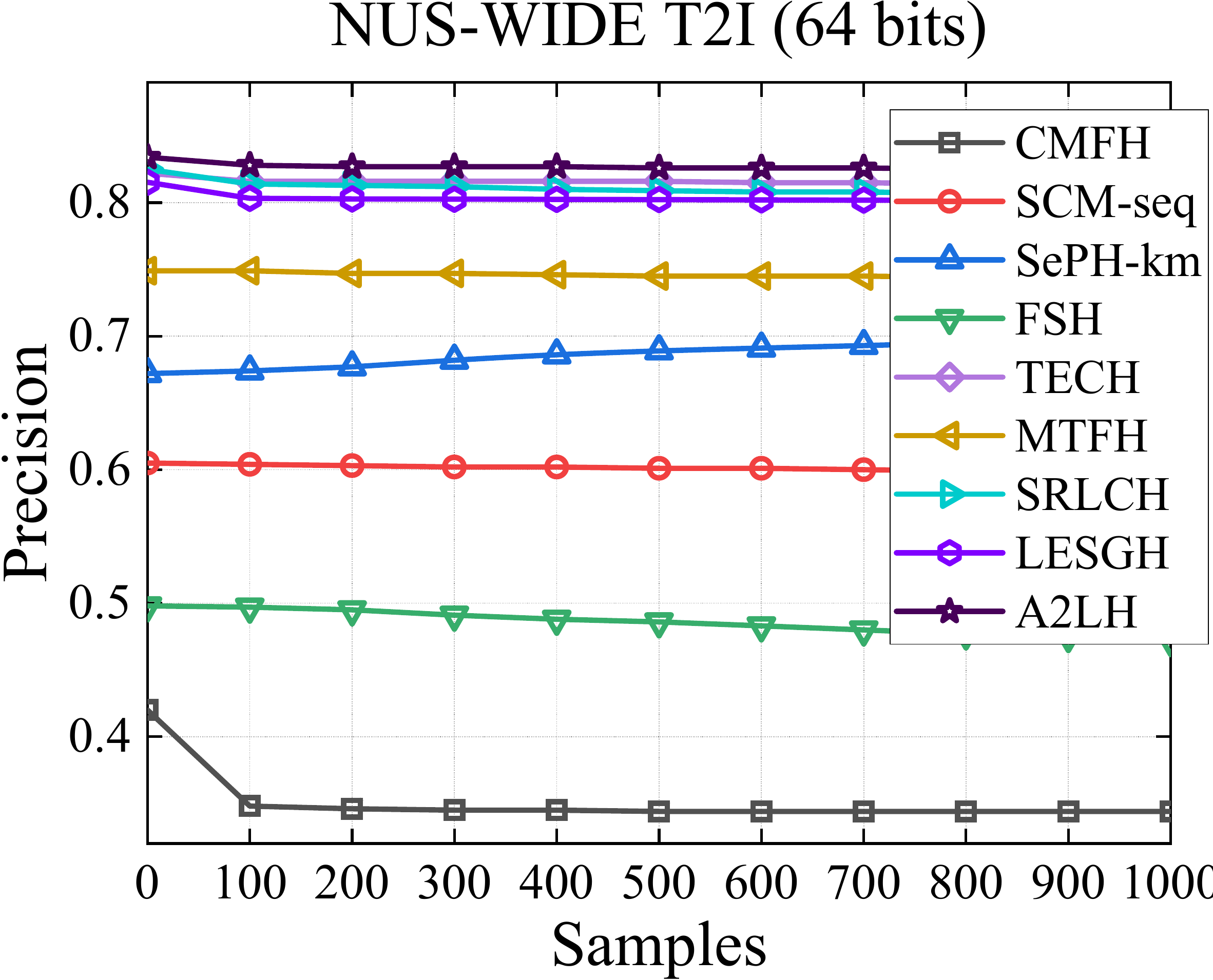}
   }
 \caption{The top-$n$ precision curves of all methods on the MIRFlickr and NUS-WIDE datasets with the code length of 64 bits.}
 \label{fig:top-n} 
 \end{figure*}

\subsection{Baselines and Implementation Details}
The state-of-the-art methods compared in this paper are broadly classified into two categories, unsupervised hashing and supervised hashing. Specifically, unsupervised hashing methods are \textbf{CMFH}~\cite{DBLP:journals/tip/DingGZG16} and \textbf{FSH}~\cite{DBLP:conf/cvpr/LiuJWHZ17} and supervised hashing methods are \textbf{SCM-seq}~\cite{DBLP:conf/aaai/ZhangL14}, \textbf{SePH-km}~\cite{DBLP:conf/cvpr/LinDH015}, \textbf{TECH}~\cite{DBLP:conf/mm/ChenWLLNX19}, \textbf{SRLCH}~\cite{DBLP:journals/tkde/ShenLYXHSH21}, \textbf{MTFH}~\cite{DBLP:journals/pami/LiuHLC21} and \textbf{LESGH}~\cite{DBLP:journals/ijon/LongSGHY22}.

The proposed \ourmodel~in this paper has the following parameters, i.e., $\zeta$, $\alpha$, $\beta$, $\eta$ and $\lambda$. The optimal parameters are set as $\zeta=2$, $\alpha=10^{-2}$, $\beta=10^1$, $\eta=10^{-3}$ and $\lambda=10^{-4}$. The number of instances selected for kernelization, i.e., $q$, is set to 2500. 

\subsection{Metrics}
In this paper, we focus on two types of retrieval tasks, i.e., retrieving texts by the use of image query (I2T) and retrieving images by the use of text query (T2I). Two broad evaluation metrics are used, i.e., mAP and P@$n$. Specifically, mAP is defined as,
\begin{equation}
mAP=\frac{1}{O_q}\frac{1}{P_q}\sum_{i=1}^{O_q}\sum_{i=1}^n Pre(i)\times \mathbb{Y}_q(i),
\end{equation}
where $O_q$ is the number of query, $P_q$ is the similar instances in the dataset, $n$ is the number of dataset, $Pre(i)$ denotes the precision of returned top $i$ instances, $\mathbb{Y}_q(i)$ represents the indicator operation, i.e., $\mathbb{Y}_q(i)=1$ means that the $i$-th returned instance is related to the query, and vice versa.

\subsection{Results}
The results of mAP are shown in Table ~\ref{tab:mAP_MIR}~\ref{tab:mAP_NUS}. From the tables, it can be observed that,
\begin{enumerate}
\item Supervised hash learning methods, i.e., CMFH and FSH, are superior to unsupervised ones, i.e., SCM-seq, SePH-km, TECH, SRLCH, MTFH, LSEGH, \ourmodel. The reason is because supervised hash learning methods consider semantic labels, which greatly enhances the discriminative power of hash code generation.
\item With the increase of hash codes, the retrieval performance is gradually improving. However, as the code bit length increases to a certain level, the retrieval performance does not improve significantly. The reason is due to the fact that as the hash code length increases, more information can be accommodated, but longer hash codes also increase noise, such as quantization error accumulation and other factors.
\item The reason why our proposed method is significantly better than the other compared state-of-the-art methods is that we incorporate both the common latent representation of different modalities and the label semantic representation into the learned hash codes to improve their discriminative power. In addition, we leverage a discrete optimization strategy to solve the problem of quantization loss caused by the relaxation-based strategy.
\end{enumerate}

In addition, we also plotted P@$n$ curves when the hash code length is 64 bits on the NUS-WIDE dataset in Fig.~\ref{fig:top-n}. Higher curves indicate better performance. From the figures, we can see that our proposed method is superior to the other methods.

\subsection{Comparison with Deep Cross-modal Hashing}
Due to the significant success of deep neural networks in the field of representation learning, and some deep cross-modal hashing methods have been proposed in the recent years. In this subsection, we present a variant of \ourmodel, i.e., \ourmodel-D, and compare it with some state-of-the-art deep cross-modal hashing methods, i.e., RDCMH~\cite{DBLP:conf/aaai/LiuYD00G19}, NrDCMH~\cite{DBLP:journals/isci/WangYZGCZ21} and DCHUC~\cite{DBLP:journals/tkde/TuMMHYWH22}. Specifically, we first extract the deep features of the image data using VGG-Net~\cite{DBLP:journals/corr/SimonyanZ14a} as image-modality representation. Thereafter, the deep 4,096d vector image modality and the original text representation for training. The mAP results are shown in Table~\ref{tab:Compare_Deep_model}. It can be seen from the table, our proposed \ourmodel-D achieves the best performance compared to the state-of-the-art deep methods, even though our proposed \ourmodel-D is not an end-to-end learning strategy. The reason is that the deep image features contain rich deep semantic information and we use a discrete optimization strategy approach to solve the quantization error problem caused by the discrete optimization-based strategy employed by the deep hashing methods. 

\begin{table*}
\centering
  \caption{The mAP values of \ourmodel-D and some deep hashing with VGG-19 features on NUS-WIDE dataset.}
  \label{tab:Compare_Deep_model}
  \begin{tabular}{l|ccc|ccc}
    \toprule
    Task & \multicolumn{3}{c|}{I$\rightarrow$T} & \multicolumn{3}{c}{T$\rightarrow$I}\cr
    \midrule
    Method & 16 bits & 32 bits & 64 bits & 16 bits & 32 bits & 64 bits\cr
    \hline
    RDCMH   & 0.6123 & 0.6213 & 0.6300 & 0.6510 & 0.6598 & 0.6357 \cr
    NrDCMH  & 0.6023 & 0.6078 & 0.6234 & 0.6432 & 0.6534 & 0.6341 \cr
    DCHUC   & 0.7335 & 0.7576 & 0.7830 & 0.6689 & 0.6995 & 0.7321\cr
    \ourmodel-D & \textbf{0.7509} & \textbf{0.7712} & \textbf{0.8023}& \textbf{0.6831} & \textbf{0.7110} & \textbf{0.7413}\cr
    \bottomrule
  \end{tabular}
\end{table*}

\subsection{Ablation Experiments}
\subsubsection{Effects of Kernelization}
The purpose of kernelization (see Sec.~\ref{sec:Kernel}) is to ensure potential nonlinear correlations within the different modalities. In order to verify the effect of kernelization operation on different modalities, in this subsection, we design a variant of \ourmodel, named \ourmodel-K, which removes the kernelization operation and perform hash code learning and hash function generation directly with the original features of different modalities. Specifically, the overall objective function~\eqref{eq:Over} can be rewritten as,
\begin{equation}\label{eq:N_kernel}
\begin{aligned}
&\underset{\mu_m,\mathbf{R,C,B,U,V}_m}{\min}~||({\mathbf{RL}})^\top\mathbf{CU}-k\mathbf{S}||_F^2+\alpha||\mathbf{B}-\mathbf{CU}||_F^2\\
&+\beta||\mathbf{RL}-\mathbf{B}||_F^2+{\mu_m}^{\zeta}||\mathbf{X}^{(m)}-{\mathbf{V}_m}\mathbf{U}||_F^2+\eta\mathcal{R}(\mathbf{RL,U})\\
&~s.t.~\mathbf{B}\in\{-1,1\}^{k\times n},~{\mathbf{V}_m}^\top\mathbf{V}_m=\mathbf{I}_q, \sum_m \mu_m=1, \mu_m>0,
\end{aligned}
\end{equation}

\subsubsection{Effects of Multi-Semantic Space Learning}
The multi-label space learning module is designed to refine the semantic association relationships between different modalities, while embedding the rich semantic information in the labels into the hash codes. In order to verify the effects of multi-label space learning module on the quality of hash code generation, in this subsection, we design a variant of \ourmodel, named \ourmodel-M, which removes the relevant module for verification. Specifically, the overall objective function~\eqref{eq:Over} can be rewritten as,
\begin{equation}\label{eq:N_MSSL}
\begin{aligned}
&\underset{\mu_m,\mathbf{R,C,B,U,V}_m}{\min}~||{\mathbf{B}}^\top\mathbf{CU}-k\mathbf{S}||_F^2+\alpha||\mathbf{B}-\mathbf{CU}||_F^2\\
&+{\mu_m}^{\zeta}||\phi(\mathbf{X}^{(m)})-{\mathbf{V}_m}\mathbf{U}||_F^2+\eta\mathcal{R}(\mathbf{U})\\
&~s.t.~\mathbf{B}\in\{-1,1\}^{k\times n},~{\mathbf{V}_m}^\top\mathbf{V}_m=\mathbf{I}_q, \sum_m \mu_m=1, \mu_m>0,
\end{aligned}
\end{equation}

\subsubsection{Effects of Discrete Optimization Scheme}
The use of discrete optimization strategy to generate hash codes can avoid the quantization errors. To verify the effectiveness of the discrete optimization strategy, we design a variant of \ourmodel, named \ourmodel-B, which first treats the hash codes $\bf B$ as real-value matrix and then uses a threshold function to process them, i.e., $\tilde{\mathbf{B}}=sgn(\mathbf{B})$. Specifically, the solution of Eq.~\eqref{eq:B_v} can be rewritten as,
\begin{equation}
\mathbf{B}=\frac{\alpha\mathbf{CU}+\beta\mathbf{RL}}{\alpha+\beta}.
\end{equation}

Table~\ref{tab:variant} reports the mAP results on the NUS-WIDE dataset with various lengths. From the table, it can be shown that (1) removing the kernelization operation leads to a degradation in the retrieval performance of hash codes due to the fact that the kernelization operation can mine the complex association relationships within different modalities and improve the generated hash code discriminative capabilities. (2) Removing the multi-semantic space learning module also leads to a decrease in retrieval performance, because the multi-semantic space learning module can improve the semantics of the generated hash codes, which in turn improves the quality of hash code generation. (3) The use of hash code optimization strategies based on relaxation strategies can lead to severe retrieval performance degradation due to the large quantization errors caused by optimization methods based on relaxation strategies.

\begin{table}
\setlength{\abovecaptionskip}{10pt}
\small
  \caption{The ablation results of \ourmodel~on the NUS-WIDE dataset with various code lengths.}
  \label{tab:variant}
  \begin{tabular}{c|l|ccccc}
    \toprule
    Task &Method & 16 bits& 32 bits&64 bits&128 bits\\
    \midrule
    \multirow{4}{0.7cm}{I$\rightarrow$T} & \ourmodel-K & 0.6268 & 0.6309 & 0.6370 & 0.6444\cr
    								      & \ourmodel-M & 0.6190 & 0.6198 & 0.6291 & 0.6404\cr
    								      & \ourmodel-B & 0.5900 & 0.6156 & 0.6189 & 0.6223\cr			     
    								      &\ourmodel & \textbf{0.6298} & \textbf{0.6388} & \textbf{0.6497} & \textbf{0.6507}\cr
    \hline
    \multirow{4}{0.7cm}{T$\rightarrow$I} & \ourmodel-K & 0.7646 & 0.7813 & 0.7887 & 0.7936\cr
    								      & \ourmodel-M & 0.7551 & 0.7752 & 0.7812 & 0.7879\cr
    								      & \ourmodel-B & 0.7328 & 0.7526 & 0.7680 & 0.7803\cr								     
    								      &\ourmodel  & \textbf{0.7698} & \textbf{0.7942} & \textbf{0.8032} & \textbf{0.8057}\cr
    \bottomrule
  \end{tabular}
\end{table}

\begin{figure}
	\centering
		\includegraphics[width=0.75\linewidth]{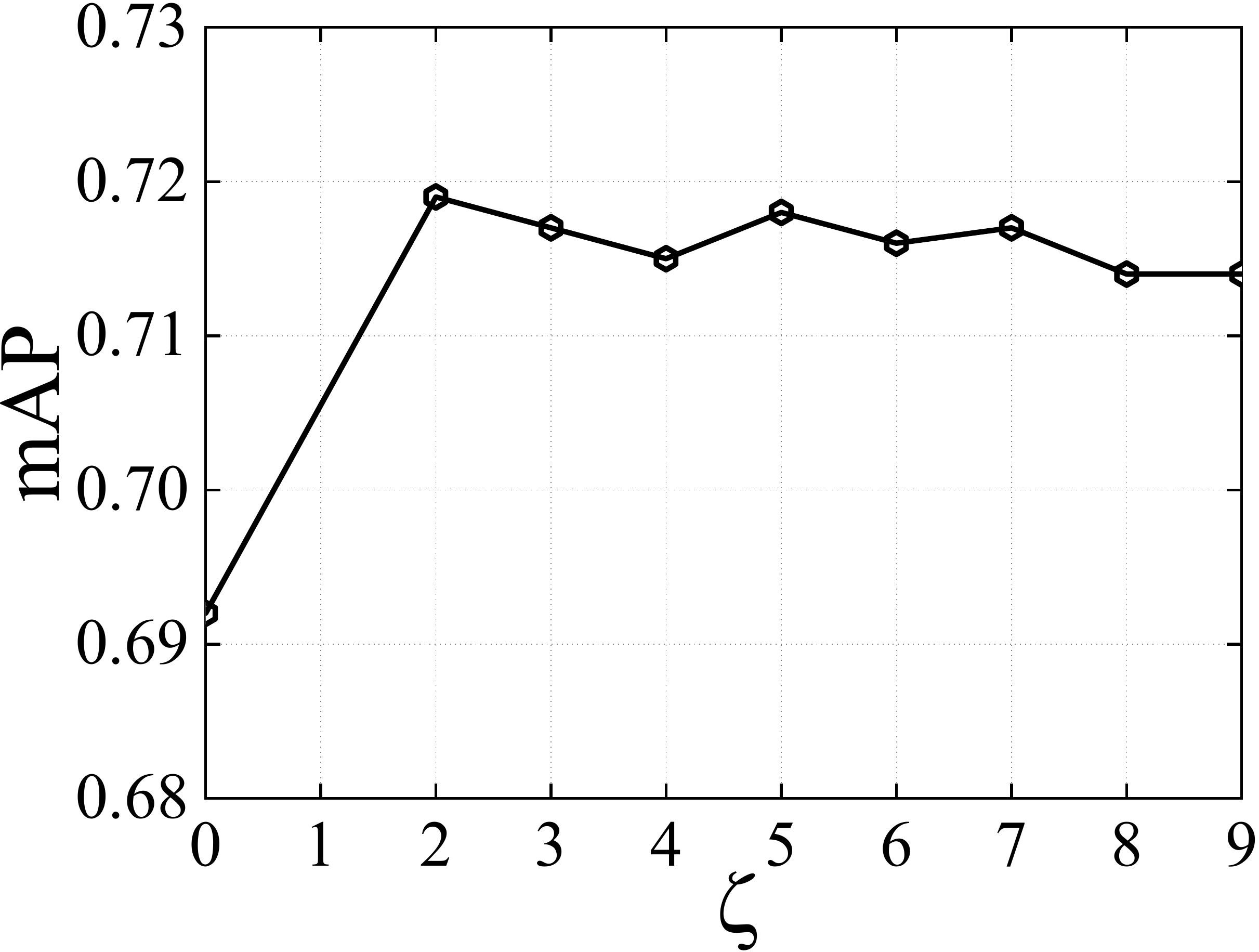}
	\caption{Parameter sensitivity analysis of $\zeta$ on the NUS-WIDE dataset@ 32 bits..}
	\label{FIG:zeta}
\end{figure}

\begin{figure}
 \centering
  \subfigure{
   \includegraphics[width=1.6in]{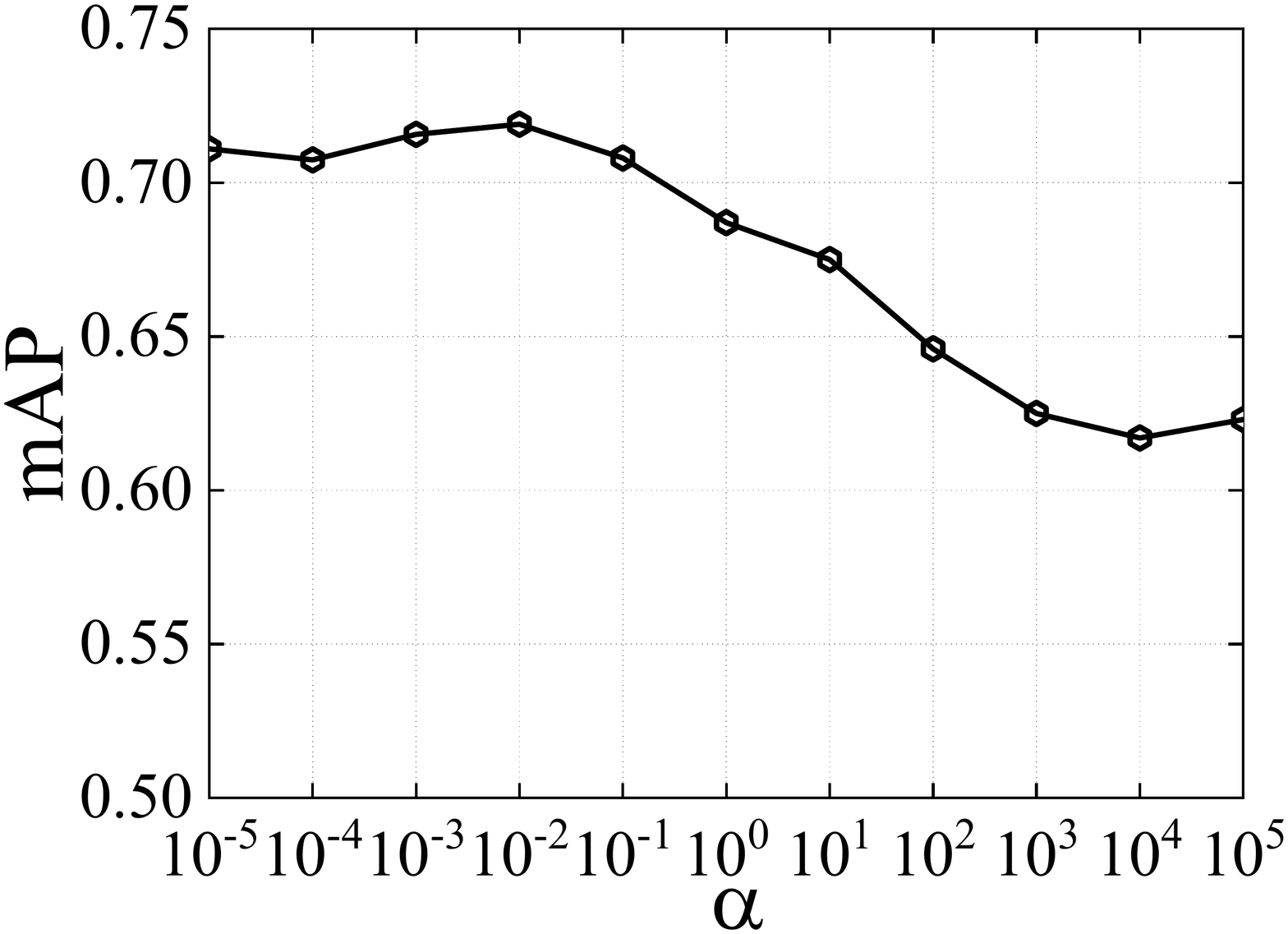}
   }
  \subfigure{
 \includegraphics[width=1.6in]{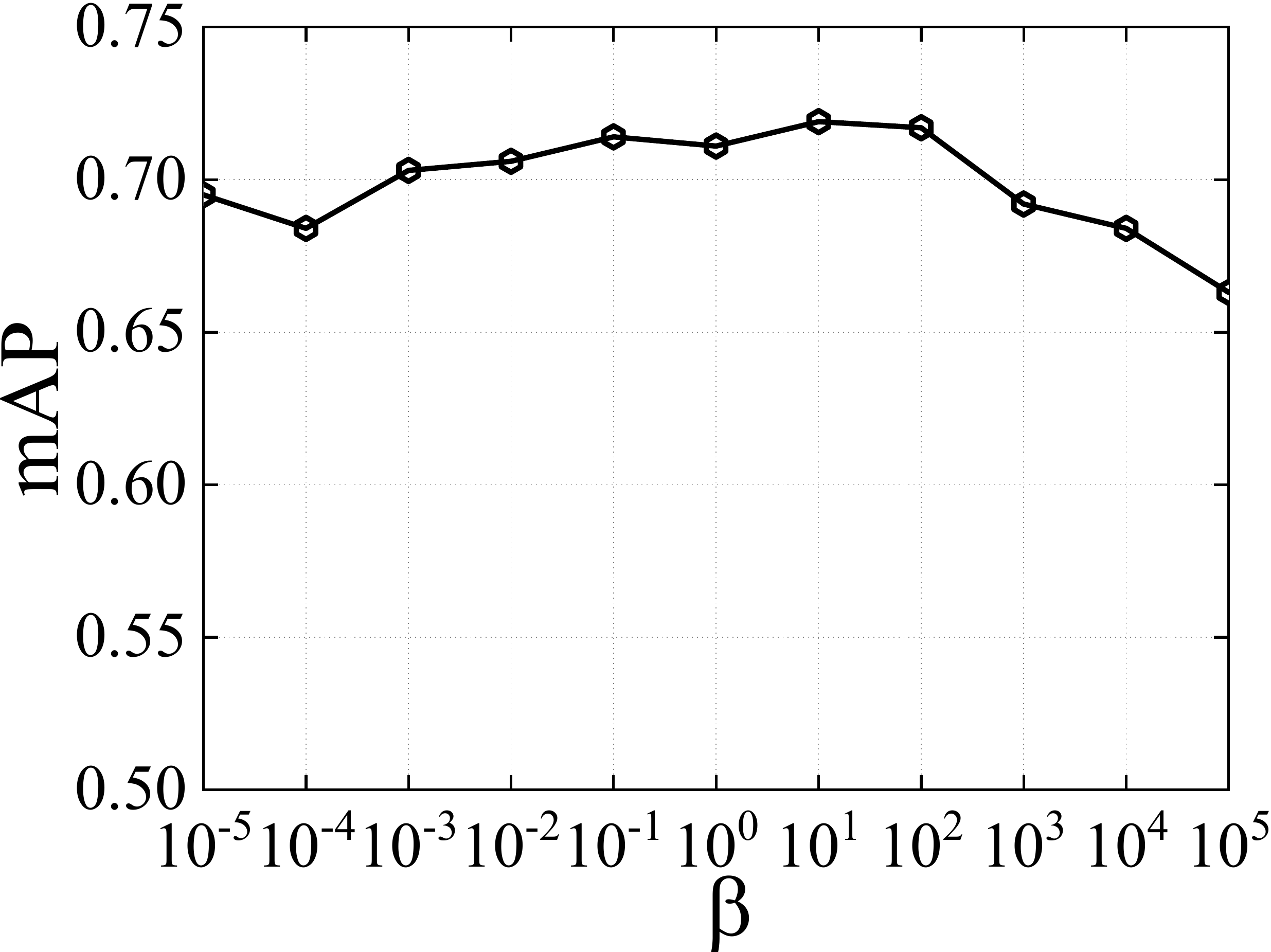}
 }
 
  \subfigure{
   \includegraphics[width=1.6in]{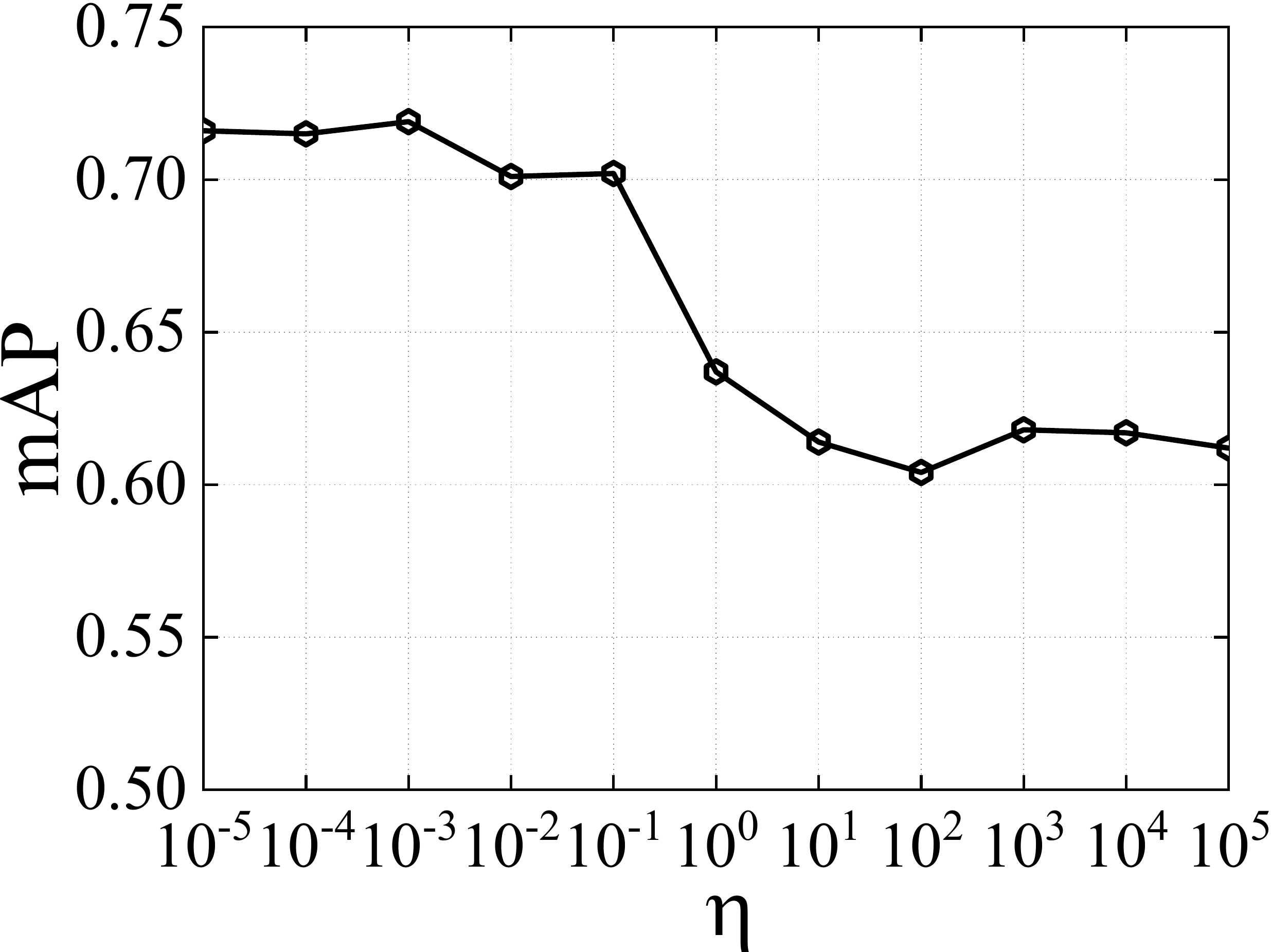}
   }
   \subfigure{
   \includegraphics[width=1.6in]{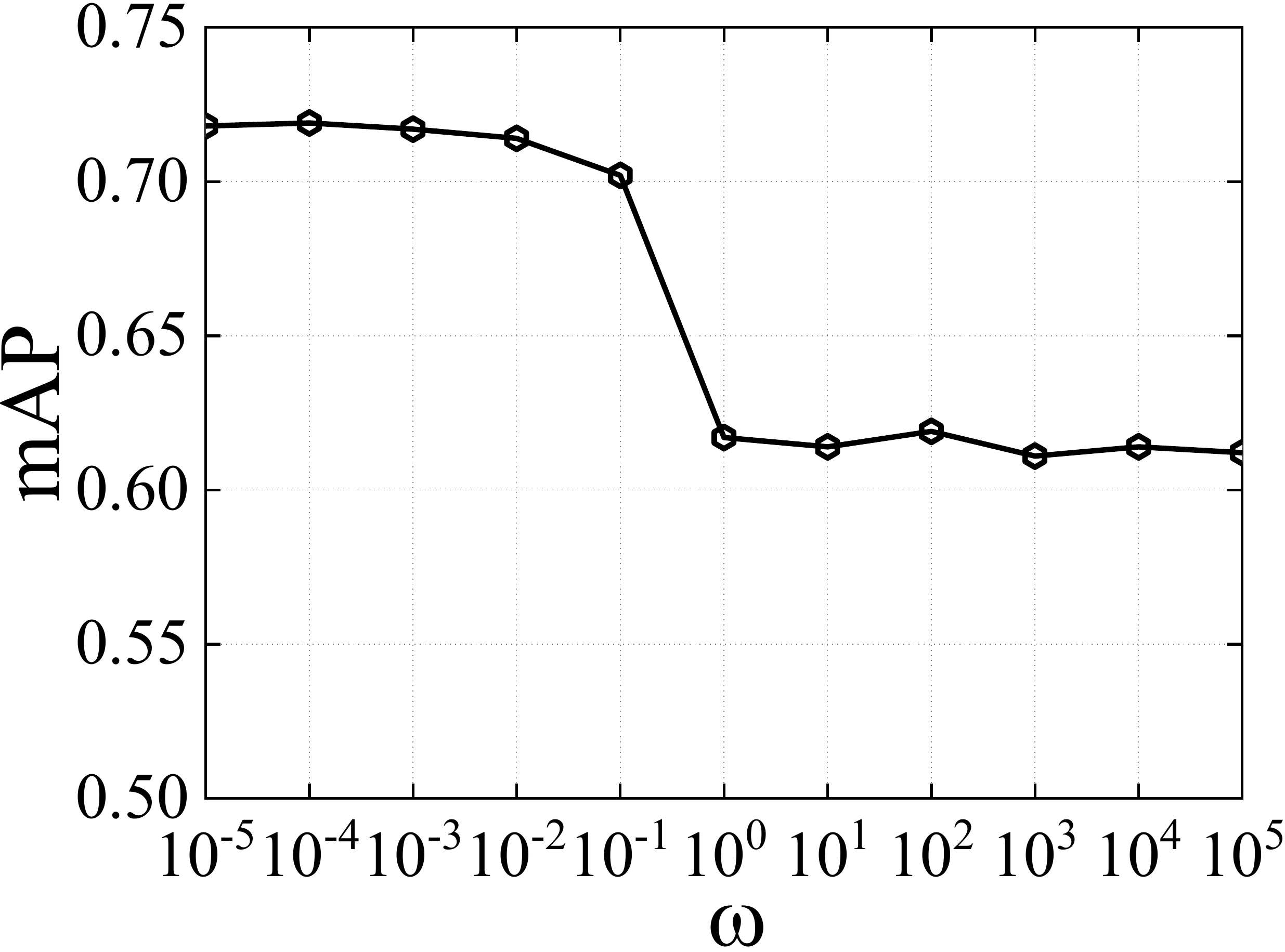}
   }
 \caption{Parameter sensitivity analysis of $\alpha,\beta,\eta,\omega$ on the NUS-WIDE dataset@ 32 bits.}
 \label{fig:Param_Sen} 
 \end{figure}

\begin{figure}
 \centering
  \subfigure{
   \includegraphics[width=1.6in]{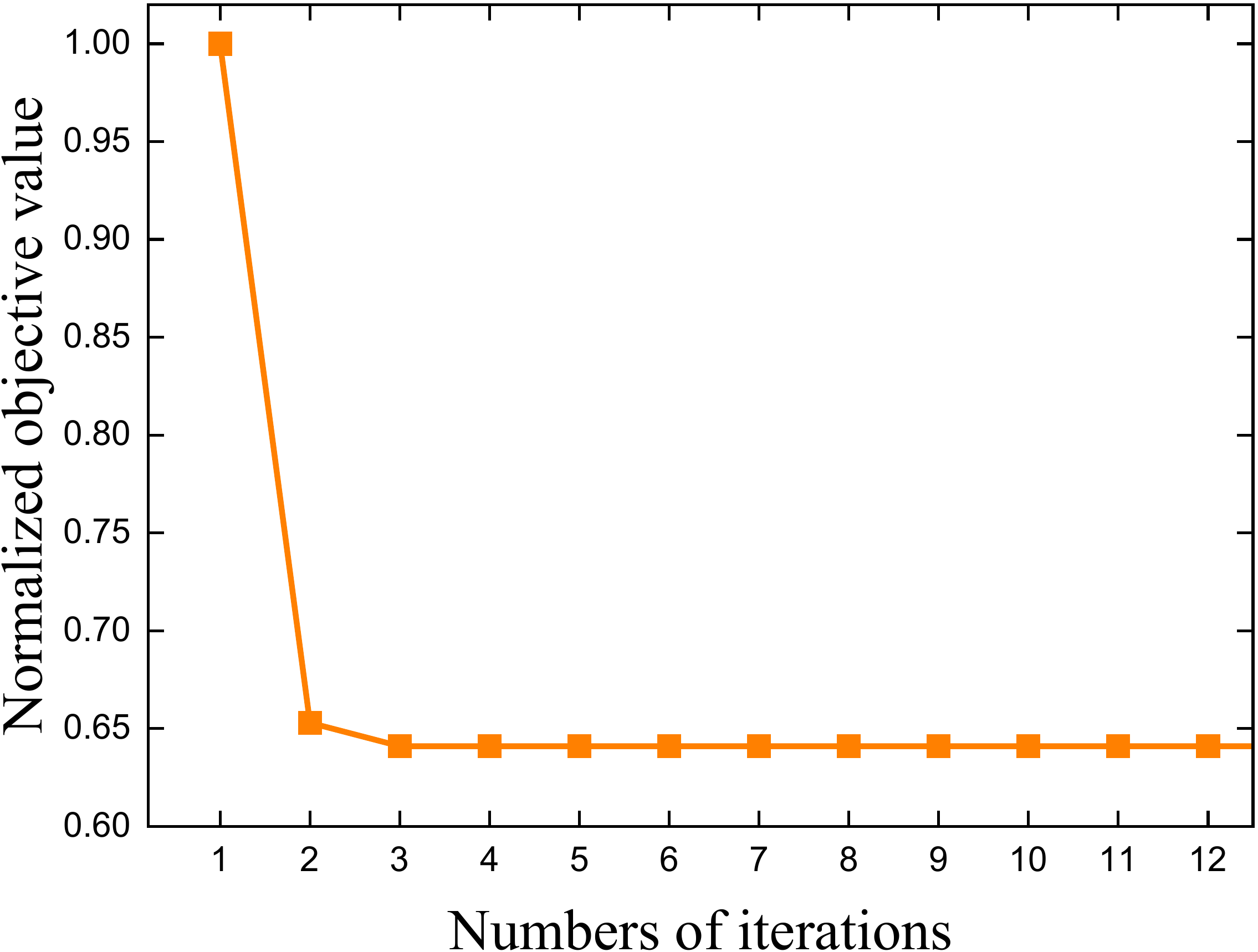}
   }
  \subfigure{
 \includegraphics[width=1.6in]{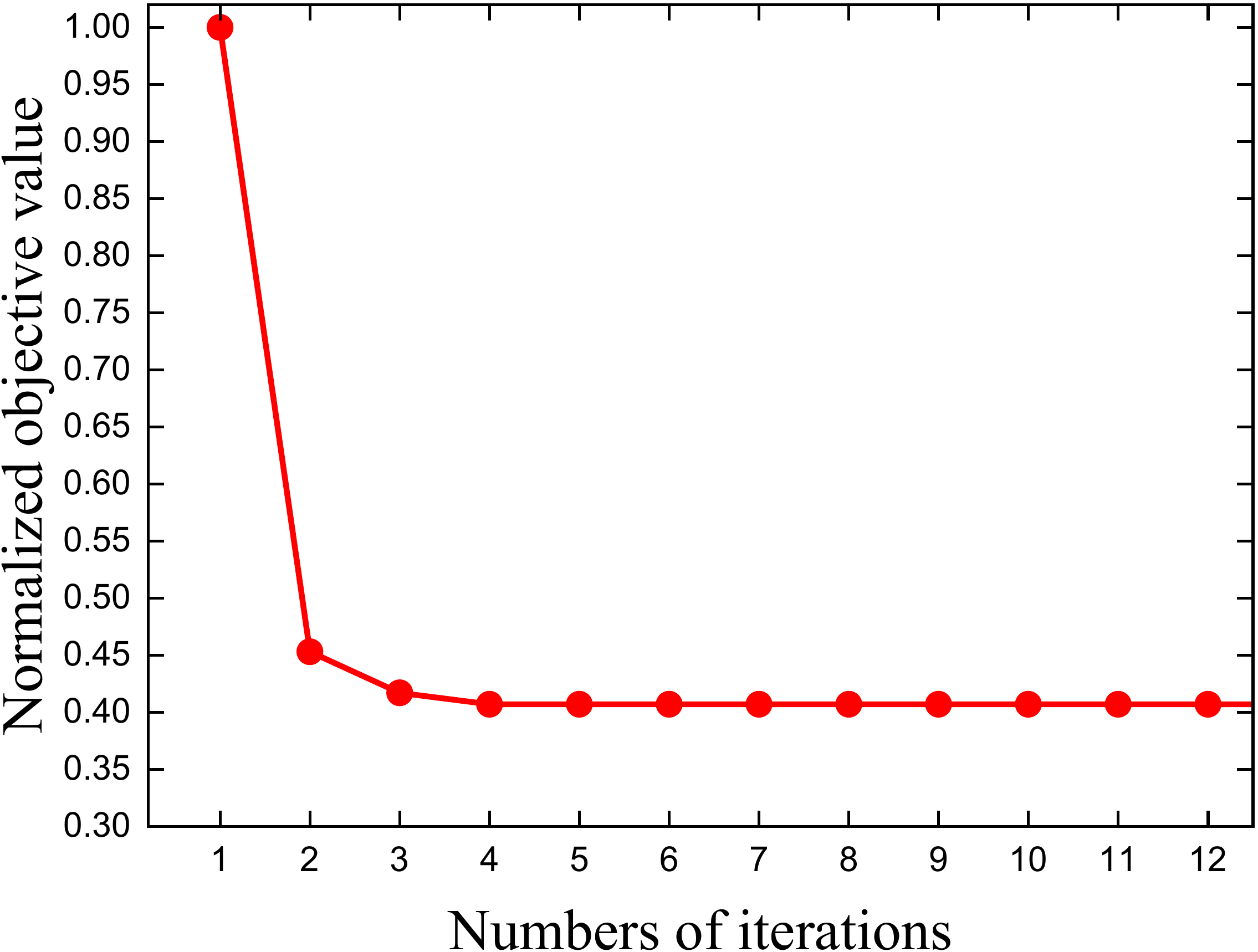}
 }
 \caption{Convergence curves of \ourmodel~on two datasets with 64 bits. \textit{\textbf{Left}}: \textit{MIRFlickr dataset}, \textit{\textbf{Right}}: \textit{NUS-WIDE dataset}.}
 \label{fig:Convergence} 
 \end{figure}

\subsection{Parameter Sensitivity Analysis}
In order to assess the impact of parameters on model performance, we performed a relevant parameter sensitivity analysis on the MIRFlickr dataset. Specifically, one of the parameters to be evaluated is updated by fixing other parameters. The length of the hash codes is set to a fixed 64 bits. The values of parameters $\alpha,\beta$, $\eta$ and $\omega$ are set in the range of $\{10^{-5},10^{-4},...,10^4,10^5\}$ and the values of the parameter $\zeta$ is set in the range of $\{0,2,3,4,...,9\}$. The mAP results are shown in Fig.~\ref{FIG:zeta} and~\ref{fig:Param_Sen}. From the figures, it can be observed that (1) the exponential parameter $\zeta$ is very stable in the ranges of $[2,9]$. It is worth noting that when $\zeta=0$, i.e., neglecting the contribution of different modalities to the common latent representation, it causes a large performance degradation. The experimental results demonstrate the effectiveness of our proposed dynamic contribution weighting strategy. (2) The parameters $\alpha,\beta$, $\eta$ and $\omega$ are very stable over a wide range of values. Specifically, the parameter $\alpha$ is stable in the range of values $[10^{-5},10^{-1}]$; the parameter $\beta$ is stable in the range of values $[10^{-5},10^{2}]$; the parameter $\eta$ is stable in the range of values $[10^{-5},10^{-3}]$; the parameter $\alpha$ is stable in the range of values $[10^{-5},10^{-1}]$. In summary, our proposed \ourmodel~can be easily deployed to practical applications.

%

\subsection{Convergence Analysis}
To verify the convergence of our proposed optimization algorithm, we will conduct convergence analysis experiments on MIRFlickr dataset. The values of the objective function~\eqref{eq:Over} are plotted in Fig.~\ref{fig:Convergence}. From the figure, it can be found that our proposed algorithm reaches convergence quickly, with less than 5 iterations, the reason is that all variables of the proposed algorithm have analytic solutions. Note that the values of objective function are normalized to between 0 and 1.


\section{Conclusion}~\label{sec:5}
This paper presents a novel discrete hashing method, termed \fullname~(\ourmodel). We design an asymmetric discrete hash learning framework that links the common latent representations of different modalities with label embeddings in a clever way, by which learning will greatly improve the discriminative power of hash codes and thus the retrieval performance. In addition, an efficient discrete optimization algorithm is proposed to solve the optimization problem of hash code learning, and the proposed method can compensate the problems of quantization error and low learning rate. Adequate experimental results show that our proposed method significantly outperforms the compared state-of-the-art methods, thus validating the effectiveness of our method. In the future, we will incorporate contrast learning techniques to improve the robustness of hash code learning.


\section*{Acknowledgment}

The authors would also like to thank the associate editor and anonymous reviewers for their comments to improve the paper.

\bibliographystyle{plain}
\bibliography{sample-base}

\begin{thebibliography}{10}

\bibitem{DBLP:journals/ipm/AnLZZL22}
Junfeng An, Haoyang Luo, Zheng Zhang, Lei Zhu, and Guangming Lu.
\newblock Cognitive multi-modal consistent hashing with flexible semantic
  transformation.
\newblock {\em Inf. Process. Manag.}, 59(1):102743, 2022.

\bibitem{DBLP:journals/tcsv/ChenLLNZX20}
Zhen{-}Duo Chen, Chuan{-}Xiang Li, Xin Luo, Liqiang Nie, Wei Zhang, and
  Xin{-}Shun Xu.
\newblock {SCRATCH:} {A} scalable discrete matrix factorization hashing
  framework for cross-modal retrieval.
\newblock {\em {IEEE} Trans. Circuits Syst. Video Technol.}, 30(7):2262--2275,
  2020.

\bibitem{DBLP:journals/tip/ChenLWGX22}
Zhen{-}Duo Chen, Xin Luo, Yongxin Wang, Shanqing Guo, and Xin{-}Shun Xu.
\newblock Fine-grained hashing with double filtering.
\newblock {\em {IEEE} Trans. Image Process.}, 31:1671--1683, 2022.

\bibitem{DBLP:conf/mm/ChenWLLNX19}
Zhen{-}Duo Chen, Yongxin Wang, Hui{-}Qiong Li, Xin Luo, Liqiang Nie, and
  Xin{-}Shun Xu.
\newblock A two-step cross-modal hashing by exploiting label correlations and
  preserving similarity in both steps.
\newblock In {\em {ACM MM}}, pages 1694--1702, 2019.

\bibitem{DBLP:journals/tip/DingGZG16}
Guiguang Ding, Yuchen Guo, Jile Zhou, and Yue Gao.
\newblock Large-scale cross-modality search via collective matrix factorization
  hashing.
\newblock {\em TIP}, 25(11):5427--5440, 2016.

\bibitem{DBLP:journals/tmm/DingFHXP17}
Kun Ding, Bin Fan, Chunlei Huo, Shiming Xiang, and Chunhong Pan.
\newblock Cross-modal hashing via rank-order preserving.
\newblock {\em {IEEE} Trans. Multimedia}, 19(3):571--585, 2017.

\bibitem{DBLP:journals/ijon/GuWZYY019}
Yifan Gu, Shidong Wang, Haofeng Zhang, Yazhou Yao, Wankou Yang, and Li~Liu.
\newblock Clustering-driven unsupervised deep hashing for image retrieval.
\newblock {\em Neurocomputing}, 368:114--123, 2019.

\bibitem{DBLP:journals/pami/GuiLSTT18}
Jie Gui, Tongliang Liu, Zhenan Sun, Dacheng Tao, and Tieniu Tan.
\newblock Fast supervised discrete hashing.
\newblock {\em {IEEE} Trans. Pattern Anal. Mach. Intell.}, 40(2):490--496,
  2018.

\bibitem{DBLP:journals/pami/HeoLHCY15}
Jae{-}Pil Heo, Youngwoon Lee, Junfeng He, Shih{-}Fu Chang, and Sung{-}Eui Yoon.
\newblock Spherical hashing: Binary code embedding with hyperspheres.
\newblock {\em {IEEE} Trans. Pattern Anal. Mach. Intell.}, 37(11):2304--2316,
  2015.

\bibitem{DBLP:journals/tip/HoangDNC20}
Tuan Hoang, Thanh{-}Toan Do, Tam~V. Nguyen, and Ngai{-}Man Cheung.
\newblock Unsupervised deep cross-modality spectral hashing.
\newblock {\em {IEEE} Trans. Image Process.}, 29:8391--8406, 2020.

\bibitem{DBLP:conf/icmi/HuangMJ19}
Jiaming Huang, Chen Min, and Liping Jing.
\newblock Unsupervised deep fusion cross-modal hashing.
\newblock In {\em International Conference on Multimodal Interaction, {ICMI}
  2019, Suzhou, China, October 14-18, 2019}, pages 358--366. {ACM}, 2019.

\bibitem{DBLP:journals/tip/JinLHQT18}
Lu~Jin, Kai Li, Hao Hu, Guo{-}Jun Qi, and Jinhui Tang.
\newblock Semantic neighbor graph hashing for multimodal retrieval.
\newblock {\em {IEEE} Trans. Image Processing}, 27(3):1405--1417, 2018.

\bibitem{DBLP:conf/aaai/LiDWXL19}
Chao Li, Cheng Deng, Lei Wang, De~Xie, and Xianglong Liu.
\newblock Coupled cyclegan: Unsupervised hashing network for cross-modal
  retrieval.
\newblock In {\em The Thirty-Third {AAAI} Conference on Artificial
  Intelligence, {AAAI} 2019, The Thirty-First Innovative Applications of
  Artificial Intelligence Conference, {IAAI} 2019, The Ninth {AAAI} Symposium
  on Educational Advances in Artificial Intelligence, {EAAI} 2019, Honolulu,
  Hawaii, USA, January 27 - February 1, 2019}, pages 176--183. {AAAI} Press,
  2019.

\bibitem{DBLP:journals/cin/LiLT0MY21}
Mingyong Li, Qiqi Li, Lirong Tang, Shuang Peng, Yan Ma, and Degang Yang.
\newblock Deep unsupervised hashing for large-scale cross-modal retrieval using
  knowledge distillation model.
\newblock {\em Comput. Intell. Neurosci.}, 2021.

\bibitem{DBLP:conf/ijcai/LiLDLG18}
Ning Li, Chao Li, Cheng Deng, Xianglong Liu, and Xinbo Gao.
\newblock Deep joint semantic-embedding hashing.
\newblock In {\em Proceedings of the Twenty-Seventh International Joint
  Conference on Artificial Intelligence, {IJCAI} 2018, July 13-19, 2018,
  Stockholm, Sweden.}, pages 2397--2403, 2018.

\bibitem{DBLP:journals/tip/LiangPLLY22}
Yuchen Liang, Yan Pan, Hanjiang Lai, Wei Liu, and Jian Yin.
\newblock Deep listwise triplet hashing for fine-grained image retrieval.
\newblock {\em {IEEE} Trans. Image Process.}, 31:949--961, 2022.

\bibitem{DBLP:journals/ijon/LinS22}
Liuyin Lin and Xin Shu.
\newblock Gaussian similarity preserving for cross-modal hashing.
\newblock {\em Neurocomputing}, 494:446--454, 2022.

\bibitem{DBLP:conf/cvpr/LinDH015}
Zijia Lin, Guiguang Ding, Mingqing Hu, and Jianmin Wang.
\newblock Semantics-preserving hashing for cross-view retrieval.
\newblock In {\em {IEEE} Conference on Computer Vision and Pattern Recognition,
  {CVPR} 2015, Boston, MA, USA, June 7-12, 2015}, pages 3864--3872. {IEEE}
  Computer Society, 2015.

\bibitem{DBLP:journals/pr/LiongLT18}
Venice~Erin Liong, Jiwen Lu, and Yap{-}Peng Tan.
\newblock Cross-modal discrete hashing.
\newblock {\em Pattern Recognit.}, 79:114--129, 2018.

\bibitem{DBLP:conf/ijcai/LiuJWH16}
Hong Liu, Rongrong Ji, Yongjian Wu, and Gang Hua.
\newblock Supervised matrix factorization for cross-modality hashing.
\newblock In {\em Proceedings of the Twenty-Fifth International Joint
  Conference on Artificial Intelligence, {IJCAI} 2016, New York, NY, USA, 9-15
  July 2016}, pages 1767--1773. {IJCAI/AAAI} Press, 2016.

\bibitem{DBLP:conf/cvpr/LiuJWHZ17}
Hong Liu, Rongrong Ji, Yongjian Wu, Feiyue Huang, and Baochang Zhang.
\newblock Cross-modality binary code learning via fusion similarity hashing.
\newblock In {\em {CVPR}}, pages 6345--6353, 2017.

\bibitem{DBLP:conf/aaai/LiuJWL16}
Hong Liu, Rongrong Ji, Yongjian Wu, and Wei Liu.
\newblock Towards optimal binary code learning via ordinal embedding.
\newblock In {\em Proceedings of the Thirtieth {AAAI} Conference on Artificial
  Intelligence, February 12-17, 2016, Phoenix, Arizona, {USA}}, pages
  1258--1265. {AAAI} Press, 2016.

\bibitem{DBLP:conf/sigir/LiuQGZY20}
Song Liu, Shengsheng Qian, Yang Guan, Jiawei Zhan, and Long Ying.
\newblock Joint-modal distribution-based similarity hashing for large-scale
  unsupervised deep cross-modal retrieval.
\newblock In {\em Proceedings of the 43rd International {ACM} {SIGIR}
  conference on research and development in Information Retrieval, {SIGIR}
  2020, Virtual Event, China, July 25-30, 2020}, pages 1379--1388. {ACM}, 2020.

\bibitem{DBLP:journals/tcyb/LiuDDLL16}
Xianglong Liu, Bowen Du, Cheng Deng, Ming Liu, and Bo~Lang.
\newblock Structure sensitive hashing with adaptive product quantization.
\newblock {\em {IEEE} Trans. Cybernetics}, 46(10):2252--2264, 2016.

\bibitem{DBLP:journals/pami/LiuHLC21}
Xin Liu, Zhikai Hu, Haibin Ling, and Yiu{-}Ming Cheung.
\newblock {MTFH:} {A} matrix tri-factorization hashing framework for efficient
  cross-modal retrieval.
\newblock {\em {IEEE} Trans. Pattern Anal. Mach. Intell.}, 43(3):964--981,
  2021.

\bibitem{DBLP:conf/aaai/LiuYD00G19}
Xuanwu Liu, Guoxian Yu, Carlotta Domeniconi, Jun Wang, Yazhou Ren, and Maozu
  Guo.
\newblock Ranking-based deep cross-modal hashing.
\newblock In {\em The Thirty-Third {AAAI} Conference on Artificial
  Intelligence, {AAAI} 2019, The Thirty-First Innovative Applications of
  Artificial Intelligence Conference, {IAAI} 2019, The Ninth {AAAI} Symposium
  on Educational Advances in Artificial Intelligence, {EAAI} 2019, Honolulu,
  Hawaii, USA, January 27 - February 1, 2019}, pages 4400--4407. {AAAI} Press,
  2019.

\bibitem{DBLP:journals/ijon/LongSGHY22}
Jun Long, Longzhi Sun, Lin Guo, Liujie Hua, and Zhan Yang.
\newblock Label embedding semantic-guided hashing.
\newblock {\em Neurocomputing}, 477:1--13, 2022.

\bibitem{DBLP:journals/tmm/MaGLHLY20}
Chao Ma, Chen Gong, Xiang Li, Xiaolin Huang, Wei Liu, and Jie Yang.
\newblock Toward making unsupervised graph hashing discriminative.
\newblock {\em {IEEE} Trans. Multimedia}, 22(3):760--774, 2020.

\bibitem{DBLP:journals/tip/MandalCB19}
Devraj Mandal, Kunal~N. Chaudhury, and Soma Biswas.
\newblock Generalized semantic preserving hashing for cross-modal retrieval.
\newblock {\em {IEEE} Trans. Image Processing}, 28(1):102--112, 2019.

\bibitem{DBLP:conf/icassp/MikriukovRD22}
Georgii Mikriukov, Mahdyar Ravanbakhsh, and Beg{\"{u}}m Demir.
\newblock Unsupervised contrastive hashing for cross-modal retrieval in remote
  sensing.
\newblock In {\em {IEEE} International Conference on Acoustics, Speech and
  Signal Processing, {ICASSP} 2022, Virtual and Singapore, 23-27 May 2022},
  pages 4463--4467. {IEEE}, 2022.

\bibitem{DBLP:conf/sigir/MoranL15}
Sean Moran and Victor Lavrenko.
\newblock Regularised cross-modal hashing.
\newblock In {\em Proceedings of the 38th International {ACM} {SIGIR}
  Conference on Research and Development in Information Retrieval, Santiago,
  Chile, August 9-13, 2015}, pages 907--910. {ACM}, 2015.

\bibitem{DBLP:journals/kbs/QiangWLXM20}
Haopeng Qiang, Yuan Wan, Ziyi Liu, Lun Xiang, and Xiaojing Meng.
\newblock Discriminative deep asymmetric supervised hashing for cross-modal
  retrieval.
\newblock {\em Knowl. Based Syst.}, 204:106188, 2020.

\bibitem{DBLP:conf/cvpr/ShenSLS15}
Fumin Shen, Chunhua Shen, Wei Liu, and Heng~Tao Shen.
\newblock Supervised discrete hashing.
\newblock In {\em {IEEE} Conference on Computer Vision and Pattern Recognition,
  {CVPR} 2015, Boston, MA, USA, June 7-12, 2015}, pages 37--45. {IEEE} Computer
  Society, 2015.

\bibitem{DBLP:journals/tkde/ShenLYXHSH21}
Heng~Tao Shen, Luchen Liu, Yang Yang, Xing Xu, Zi~Huang, Fumin Shen, and
  Richang Hong.
\newblock Exploiting subspace relation in semantic labels for cross-modal
  hashing.
\newblock {\em {IEEE} Trans. Knowl. Data Eng.}, 33(10):3351--3365, 2021.

\bibitem{DBLP:journals/corr/abs-2202-10232}
Yang Shi and Young{-}joo Chung.
\newblock Efficient cross-modal retrieval via deep binary hashing and
  quantization.
\newblock {\em CoRR}, abs/2202.10232, 2022.

\bibitem{DBLP:journals/tip/ShiNLZY22}
Yang Shi, Xiushan Nie, Xingbo Liu, Li~Zou, and Yilong Yin.
\newblock Supervised adaptive similarity matrix hashing.
\newblock {\em {IEEE} Trans. Image Process.}, 31:2755--2766, 2022.

\bibitem{DBLP:journals/corr/SimonyanZ14a}
Karen Simonyan and Andrew Zisserman.
\newblock Very deep convolutional networks for large-scale image recognition.
\newblock In {\em 3rd International Conference on Learning Representations},
  2015.

\bibitem{DBLP:conf/iccv/SuZZ19}
Shupeng Su, Zhisheng Zhong, and Chao Zhang.
\newblock Deep joint-semantics reconstructing hashing for large-scale
  unsupervised cross-modal retrieval.
\newblock In {\em 2019 {IEEE/CVF} International Conference on Computer Vision,
  {ICCV} 2019, Seoul, Korea (South), October 27 - November 2, 2019}, pages
  3027--3035. {IEEE}, 2019.

\bibitem{DBLP:journals/tip/TangWS16}
Jun Tang, Ke~Wang, and Ling Shao.
\newblock Supervised matrix factorization hashing for cross-modal retrieval.
\newblock {\em {IEEE} Trans. Image Process.}, 25(7):3157--3166, 2016.

\bibitem{DBLP:journals/tkde/TuMMHYWH22}
Rong{-}Cheng Tu, Xian{-}Ling Mao, Bing Ma, Yong Hu, Tan Yan, Wei Wei, and Heyan
  Huang.
\newblock Deep cross-modal hashing with hashing functions and unified hash
  codes jointly learning.
\newblock {\em {IEEE} Trans. Knowl. Data Eng.}, 34(2):560--572, 2022.

\bibitem{DBLP:journals/pr/WangWHGT20}
Di~Wang, Quan Wang, Lihuo He, Xinbo Gao, and Yumin Tian.
\newblock Joint and individual matrix factorization hashing for large-scale
  cross-modal retrieval.
\newblock {\em Pattern Recognit.}, 107:107479, 2020.

\bibitem{DBLP:journals/pami/WangZSSS18}
Jingdong Wang, Ting Zhang, Jingkuan Song, Nicu Sebe, and Heng~Tao Shen.
\newblock A survey on learning to hash.
\newblock {\em {IEEE} Trans. Pattern Anal. Mach. Intell.}, 40(4):769--790,
  2018.

\bibitem{DBLP:journals/isci/WangYZGCZ21}
Runmin Wang, Guoxian Yu, Hong Zhang, Maozu Guo, Lizhen Cui, and Xiangliang
  Zhang.
\newblock Noise-robust deep cross-modal hashing.
\newblock {\em Inf. Sci.}, 581:136--154, 2021.

\bibitem{DBLP:journals/ijon/WangZCLG20}
Tong Wang, Lei Zhu, Zhiyong Cheng, Jingjing Li, and Zan Gao.
\newblock Unsupervised deep cross-modal hashing with virtual label regression.
\newblock {\em Neurocomputing}, 386:84--96, 2020.

\bibitem{DBLP:conf/sigir/WangLWZZ15}
Yang Wang, Xuemin Lin, Lin Wu, Wenjie Zhang, and Qing Zhang.
\newblock {LBMCH:} learning bridging mapping for cross-modal hashing.
\newblock In {\em Proceedings of the 38th International {ACM} {SIGIR}
  Conference on Research and Development in Information Retrieval, Santiago,
  Chile, August 9-13, 2015}, pages 999--1002. {ACM}, 2015.

\bibitem{DBLP:conf/ijcai/WuLHLDZS18}
Gengshen Wu, Zijia Lin, Jungong Han, Li~Liu, Guiguang Ding, Baochang Zhang, and
  Jialie Shen.
\newblock Unsupervised deep hashing via binary latent factor models for
  large-scale cross-modal retrieval.
\newblock In {\em Proceedings of the Twenty-Seventh International Joint
  Conference on Artificial Intelligence, {IJCAI} 2018, July 13-19, 2018,
  Stockholm, Sweden}, pages 2854--2860. ijcai.org, 2018.

\bibitem{DBLP:journals/mta/XieZC16}
Liang Xie, Lei Zhu, and Guoqi Chen.
\newblock Unsupervised multi-graph cross-modal hashing for large-scale
  multimedia retrieval.
\newblock {\em Multim. Tools Appl.}, 75(15):9185--9204, 2016.

\bibitem{DBLP:conf/sigir/YangL0H20}
Zhan Yang, Jun Long, Lei Zhu, and Wenti Huang.
\newblock Nonlinear robust discrete hashing for cross-modal retrieval.
\newblock In {\em Proceedings of the 43rd International {ACM} {SIGIR}
  conference on research and development in Information Retrieval, {SIGIR}
  2020, Virtual Event, China, July 25-30, 2020}, pages 1349--1358. {ACM}, 2020.

\bibitem{DBLP:journals/ijon/YangRHLZL20}
Zhan Yang, Osolo~Ian Raymond, Wenti Huang, Zhifang Liao, Lei Zhu, and Jun Long.
\newblock Scalable deep asymmetric hashing via unequal-dimensional embeddings
  for image similarity search.
\newblock {\em Neurocomputing}, 412:262--275, 2020.

\bibitem{DBLP:journals/ipm/YangYHSL21}
Zhan Yang, Liu Yang, Wenti Huang, Longzhi Sun, and Jun Long.
\newblock Enhanced deep discrete hashing with semantic-visual similarity for
  image retrieval.
\newblock {\em Inf. Process. Manag.}, 58(5):102648, 2021.

\bibitem{DBLP:journals/kbs/YangYRZHLL21}
Zhan Yang, Liu Yang, Osolo~Ian Raymond, Lei Zhu, Wenti Huang, Zhifang Liao, and
  Jun Long.
\newblock {NSDH:} {A} nonlinear supervised discrete hashing framework for
  large-scale cross-modal retrieval.
\newblock {\em Knowl. Based Syst.}, 217:106818, 2021.

\bibitem{DBLP:journals/cogcom/YuWZ22}
Jun Yu, Xiao{-}Jun Wu, and Donglin Zhang.
\newblock Unsupervised multi-modal hashing for cross-modal retrieval.
\newblock {\em Cogn. Comput.}, 14(3):1159--1171, 2022.

\bibitem{DBLP:journals/access/YuanDH17}
Tongtong Yuan, Weihong Deng, and Jiani Hu.
\newblock Distortion minimization hashing.
\newblock {\em {IEEE} Access}, 5:23425--23435, 2017.

\bibitem{DBLP:conf/aaai/ZhangL14}
Dongqing Zhang and Wu{-}Jun Li.
\newblock Large-scale supervised multimodal hashing with semantic correlation
  maximization.
\newblock In {\em Proceedings of the Twenty-Eighth {AAAI} Conference on
  Artificial Intelligence, July 27 -31, 2014, Qu{\'{e}}bec City, Qu{\'{e}}bec,
  Canada}, pages 2177--2183. {AAAI} Press, 2014.

\bibitem{DBLP:conf/aaai/ZhangXLLH19}
Zheng Zhang, Guo{-}Sen Xie, Yang Li, Sheng Li, and Zi~Huang.
\newblock {SADIH:} semantic-aware discrete hashing.
\newblock In {\em The Thirty-Third Conference on Artificial Intelligence},
  pages 5853--5860, 2019.

\bibitem{DBLP:journals/tmm/ZhangZLCW20}
Zheng Zhang, Qin Zou, Yuewei Lin, Long Chen, and Song Wang.
\newblock Improved deep hashing with soft pairwise similarity for multi-label
  image retrieval.
\newblock {\em {IEEE} Trans. Multimedia}, 22(2):540--553, 2020.

\bibitem{DBLP:journals/pami/ZhouYWLLT16}
Wengang Zhou, Ming Yang, Xiaoyu Wang, Houqiang Li, Yuanqing Lin, and Qi~Tian.
\newblock Scalable feature matching by dual cascaded scalar quantization for
  image retrieval.
\newblock {\em {IEEE} Trans. Pattern Anal. Mach. Intell.}, 38(1):159--171,
  2016.

\bibitem{DBLP:conf/cvpr/ZhuangLSR16}
Bohan Zhuang, Guosheng Lin, Chunhua Shen, and Ian~D. Reid.
\newblock Fast training of triplet-based deep binary embedding networks.
\newblock In {\em 2016 {IEEE} Conference on Computer Vision and Pattern
  Recognition, {CVPR} 2016, Las Vegas, NV, USA, June 27-30, 2016}, pages
  5955--5964, 2016.

\bibitem{DBLP:conf/mir/ZhuoLHHL22}
Yaoxin Zhuo, Yikang Li, Jenhao Hsiao, Chiuman Ho, and Baoxin Li.
\newblock Clip4hashing: Unsupervised deep hashing for cross-modal video-text
  retrieval.
\newblock In {\em {ICMR} '22: International Conference on Multimedia Retrieval,
  Newark, NJ, USA, June 27 - 30, 2022}, pages 158--166. {ACM}, 2022.

\end{thebibliography}


\end{document}